\documentclass[graybox]{svmult}

\usepackage{amsmath}
\usepackage{bbold}
\usepackage{mathptmx}       
\usepackage{helvet}         
\usepackage{courier}        
\usepackage{type1cm}        
\usepackage{makeidx}         
\usepackage{graphicx}        
\usepackage{multicol}        
\usepackage[bottom]{footmisc}
\usepackage{graphicx}
\usepackage{amssymb}
\usepackage{epstopdf}
\usepackage{bm}
\usepackage{graphicx}
\usepackage{color}
\usepackage[normalem]{ulem}

\makeindex             

\begin{document}

\title*{Continuum models of collective cell migration}
\author{Shiladitya Banerjee and M. Cristina Marchetti}
\institute{Shiladitya Banerjee \at University College London, London, UK, \email{shiladitya.banerjee@ucl.ac.uk}
\and M. Cristina Marchetti \at University of California Santa Barbara, Santa Barbara, CA, USA, \email{cmarchetti@ucsb.edu}}
%
%
\maketitle

\abstract{Collective cell migration plays a central role in tissue development, morphogenesis, wound repair and cancer progression. With the growing realization that physical forces mediate cell motility in development and physiology, a key biological question is how cells integrate molecular activities for force generation on multicellular scales. In this review we discuss recent advances in modeling collective cell migration using quantitative tools and approaches rooted in soft matter physics. We focus on theoretical models of cell aggregates as continuous active media, where the feedback between mechanical forces and regulatory biochemistry gives rise to rich collective dynamical behavior. This class of models provides a powerful predictive framework for the physiological dynamics that underlies many developmental processes, where cells need to collectively migrate like a viscous fluid to reach a target region, and then stiffen to support mechanical stresses and maintain tissue cohesion.\\}


\section{Introduction}
\label{sec:1}
In many physiological and developmental contexts, groups of cells coordinate their behavior to organize in coherent structures or migrate collectively~\cite{friedl2009}. Many experimental studies have established that these multicellular processes are regulated by the cross-talk between cell-cell adhesions, cell interaction with the extracellular matrix, and myosin-based contractility of the cell cortex~\cite{ladoux2017}. Importantly, faithful execution of multicellular processes requires both biochemical signaling and mechanical force transmission.

A well-studied multicellular process is wound healing, where epithelial cells march in unison to fill in a gap in the tissue~\cite{fenteany2000,begnaud2016}. 
Although the cells at the front of the advancing monolayer often show large, spread-out lamellipodia and an almost mesenchymal phenotype, long-range collective migration is not simply achieved via the pulling action of such {\it leader} cells on a sheet of inert followers~\cite{trepat2009}. In fact, traction forces transmitted to the extracellular matrix are found to remain significant well behind the leading edge of the tissue, indicating that cells in the bulk participate in force generation and transmission. 
This observation, together with the presence of spread-out cells with large cryptic lamellipodia throughout the monolayer~\cite{farooqui2005}, indicates that, although leader cells at the sheet edge provide guidance for migration, they do not play a unique role in force generation. Instead,  
a new paradigm has emerged where collective migration is associated with long-range forces extending throughout the tissue, with waves of propagating mechanical stress that are sustained by biochemical signaling at the molecular scale~\cite{trepat2009,serra2012}. These waves of stress and cellular deformation provide a mechanism for information transmission, much like sound in air. Such mechanical waves have been shown to drive periodic cycles of effective stiffening and fluidification in expanding cell monolayers~\cite{serra2012} and coherent vortical or standing motions in confined ones~\cite{doxzen2013,deforet2014}.

Multicellularity and collective migration is intimately related to the materials  properties of tissues - viscoelastic materials with both  fluid and solid-like behavior. In morphogenesis, for instance, cells must sort and flow like a liquid to reach the right location, but then stiffen and support mechanical stresses once the tissue has achieved the desired structure~\cite{lecuit2011}. 
Recent experiments have suggested that dense tissues may be in a {\it glassy} or {\it jammed} state, where local cell rearrangements are rare and energetically costly. A relatively small change in tissue mechanical parameters may trigger a change from an elastic response to viscous fluid-like behavior, where individual cells are highly motile and rearrange continuously~\cite{angelini2010, angelini2011}. 
Indeed living tissues appear to have well-defined mechanical properties, some familiar from conventional matter, such as  elastic moduli~\cite{discher2005} and surface tension~\cite{foty1994}, others unique to living systems, such as {\it homeostatic pressure}, proposed theoretically as a factor controlling tumor growth~\cite{basan2009}. 

Just like intermolecular forces yield the emergence of  materials properties in nonliving matter,  cell-cell interactions, mediated by cadherins, play a crucial role in controlling the macroscopic properties of groups of cells and tissues~\cite{mertz2013,maruthamuthu2011}. The collective mechanics of living matter, however, is more complex than that of inert materials as individual cell activity competes with cell-cell interactions in controlling the large scale behavior of cell assemblies. In addition, physical models of collective cell behavior must also incorporate interactions of cells with the extracellular matrix.
In other words, the coupling of of cells to their surroundings is affected by intracellular contractility and cell-cell interactions, which in turn can be actively regulated by the environment, in a complex feedback loop unique to living matter. Finally, unlike inert materials where phase changes are controlled by externally tuning parameters such as temperature and density, living matter can tune itself between states with different macroscopic properties through the regulation of molecular scale and genetic processes that drive motility, division, death and phenotypical changes. A quantitative understanding of the relative importance of mechanical and biochemical mechanisms in controlling the collective tissue properties is beginning to emerge through developments in molecular biology, microscopy, super-resolution imaging and force measurement techniques~\cite{roca2017}. These advances provide an ideal platform for constructing quantitative physical models that account for the role of active cellular processes in controlling collective mechanics of motile and deformable multicellular structures.

Theoretical modeling of multicellular processes can be divided broadly into two classes. The first encompasses discrete mesoscale models that incorporate some minimal features of individual cells, such as contractility and motility, and then examine how cell-cell interactions and coupling to the environment determine materials properties at the tissue scale. This class includes models of cells as active particles endowed with persistent motility~\cite{basan2013,camley2017}, as well as models that have been used extensively in developmental biology, such as Vertex~\cite{honda1980,fletcher2014}, Voronoi~\cite{li2014,bi2016} and Cellular Potts models~\cite{graner1992} that are designed to capture the behavior of confluent tissues, where there are no gaps nor overlaps between cells. Vertex and Voronoi models describe cells as irregular polygons tiling the plane and are defined by an energy functional that tends to adjust the area and perimeter of each cell to target values~\cite{farhadifar2007}. Recent modifications have also endowed these mesoscopic models with cell motility~\cite{bi2016,barton2017,staddon2018} and active contractility~\cite{noll2017}. Vertex models have been employed successfully to quantify how intercellular forces control shape at both the cell and tissue scale under the assumption of force balance at every vertex of the cellular network~\cite{farhadifar2007}. An active version of the Voronoi model was recently shown to exhibit a liquid-solid transition of confluent epithelia tuned by motility and cell shape, which in turns encodes information about the interplay between cortex contractility and cell-cell adhesion~\cite{bi2016}. An intriguing prediction of this work is that individual cell shape, that can be inferred directly from cell imaging segmentation, provides a measure of tissue rigidity~\cite{bi2015}.

The second class of theoretical work encompasses continuum models, such as phase field~\cite{ziebert2011} and active gel models~\cite{prost2015}, where a cell sheet is described as a fluid or an elastic continuum, with couplings to internal degrees of freedom that account for active processes, such as contractiity and cellular polarization. Continuum models have been shown to account for the heterogeneous spatial distribution of cellular stresses inferred from Traction Force Microscopy~\cite{style2014} in both expanding~\cite{trepat2009,serra2012,banerjee2011,blanch2017} and confined monolayers~\cite{notbohm2016}, and even at the level of individual cells~\cite{oakes2014}. They also capture the mechanical waves observed in these systems~\cite{serra2012,banerjee2015}. This review does not aim to be comprehensive, and will focus on models of tissue as active continuous media, with an emphasis on models that describe tissue as active \textit{elastic} continua. This class of mechanochemical models has had a number of successes in capturing the tissue scale behavior in adherent~\cite{mertz2012}, confined~\cite{notbohm2016} and expanding epithelia~\cite{banerjee2015}. 

Both the mesoscale and continuum approaches do not attempt to faithfully incorporate intracellular processes, but rather aim at characterizing quantitatively the modes of organization and the materials properties of cell collectives in terms of a few macroscopic parameters, such as cell density and shape, cell-cell adhesiveness, contractility, polarization and division/death rates. Each of these quantities may describe the combined effect of a number of molecular processes and signaling pathways. This approach, inspired from condensed matter physics~\cite{marchetti2013}, aims at providing experimentalists with testable predictions that may allow to correlate classes of signaling pathways to tissue scale organization.

The review is organized as follows. In Sect.~\ref{sec:2} we describe a dynamical model of cell collectives as active viscoelastic media, coupled to the dynamics of active intracellular processes such as actomyosin contractility and cell polarization. An important aspect of the model is a dynamic feedback between mechanical stresses and regulatory biochemistry which gives rise to rich collective behavior. In Sect.~\ref{sec:3} we discuss applications of this class of continuum models to describing force transmission in epithelial monolayers, waves in expanding cell sheets, collective cell migration in confinement and during epithelial gap closure. We then compare the quantitative predictions of viscoelastic solid models with fluid models of tissues in Sect.~\ref{sec:4}, describing their equivalence as well as highlighting the key differences. We conclude with a critical discussion of the continuum model limitations and highlight open theoretical questions in understanding the collective behavior of multicellular assemblies (Sect.~\ref{sec:5}). 

\section{Cells as active continuous media}\label{sec:2}
\begin{figure}[t]
\sidecaption
\includegraphics[scale=.65]{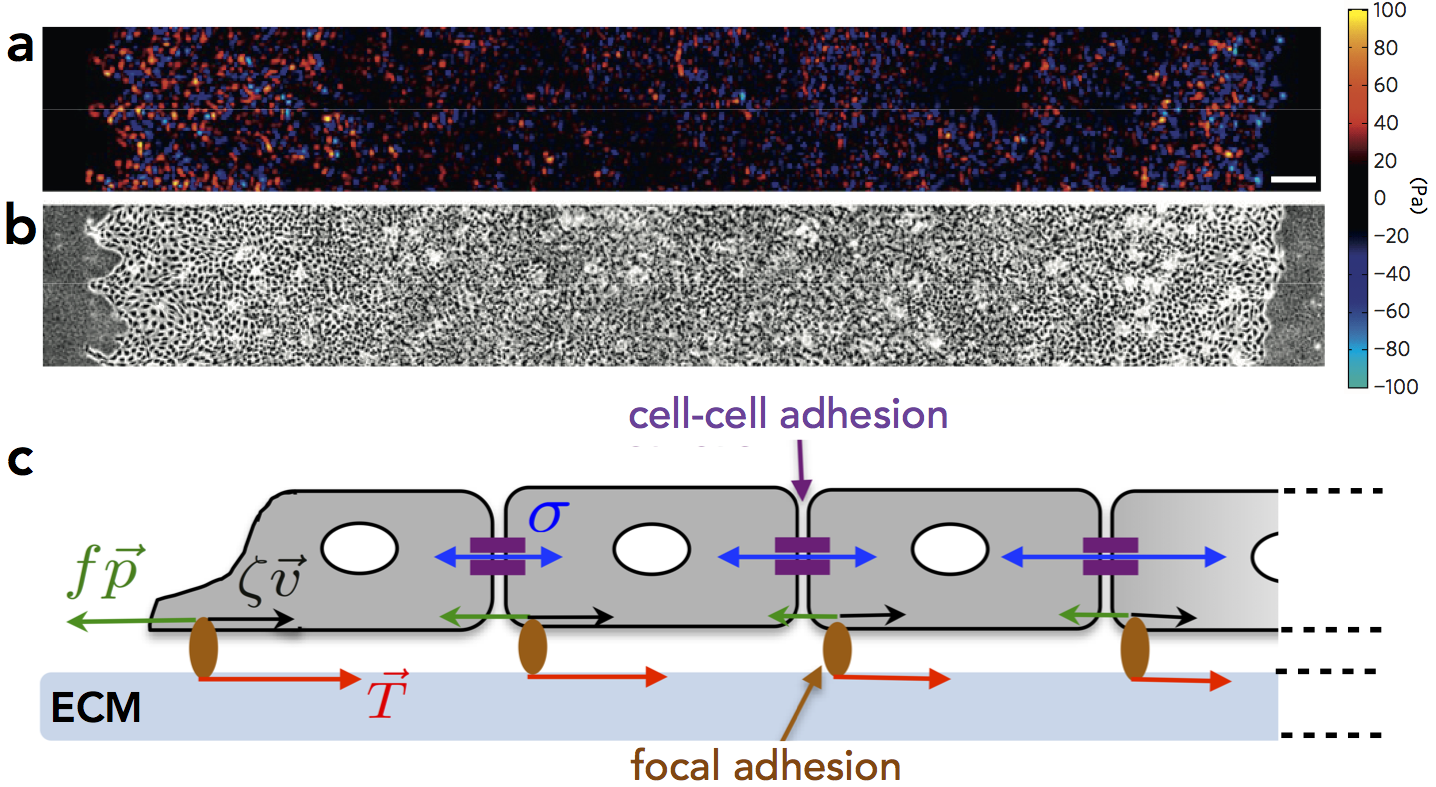}
\caption{{\bf Forces driving collective cell motion}. (a-b) Radial component of traction stress (a) and phase contrast images of an expanding MDCK cell monolayer, reproduced from Ref.~\cite{trepat2009} (scale bar=200 $\mu$m). (b) Schematic of the physical forces acting on the cell monolayer~\cite{notbohm2016}. Tractions exerted by the monolayer on the substrate (ECM) point inward (red arrows) at the monolayer edge and balance the forces due to viscous friction, $\zeta {\bf v}$ (black arrows), and polarized motility, $f{\bf p}$ (green arrows). The tractions are locally balanced by the divergence of the monolayer stress, ${\bf T}=h\nabla.\bm{\sigma}$. }
\label{fig:1}       
\end{figure}

We begin by considering the mechanics of a monolayer of epithelial cells, migrating on a soft elastic matrix (Fig.~\ref{fig:1}a-b), with an average height $h$ much thinner than in-plane cell dimensions~\cite{banerjee2011,banerjee2012,schwarz2013}. In mechanical equilibrium, the condition of local force-balance translates to $\partial_\beta \Sigma_{\alpha \beta}=0$, where ${\bm \Sigma}$ is the three-dimensional stress tensor of the monolayer, with greek indices taking values $x,y$ and $z$. In-plane force balance is given by
\begin{equation}
\label{eq:balance-2d}
\partial_j\Sigma_{ij} + \partial_z \Sigma_{iz}=0\;,
\end{equation}
with $i, j$ denoting in-plane coordinates. For a thin cell monolayer we average the cellular force-balance equation over the cell thickness $h$. We assume that the top surface of the cell is stress free, $\Sigma_{iz}({\bf r}_\perp,z=h)=0$, whereas at the cell-substrate interface, $z=0$, the cells experience lateral traction stresses given by $\Sigma_{iz}({\bf r}_\perp,z=0)=T_i({\bf r}_\perp)$. A representative traction stress map for a monolayer expanding in free space is reproduced in Fig.~\ref{fig:1}b, which shows appreciable traction stress penentration throughout bulk of the tissue. The thickness-averaged force balance equation then reads,
\begin{equation}\label{eq:force-balance}
h\partial_j\sigma_{ij}=T_i\;,
\end{equation}
where $\sigma_{ij}({\bf r}_\perp)=\int_0^h(dz/h)\Sigma_{ij}({\bf r}_\perp,z)$ is the in-plane monolayer stress. The force-balance diagram is illustrated in Fig.~\ref{fig:1}c. It is worthwhile to mention that the assumption of in-plane traction forces is a good approximation for fully spread cells making almost zero contact angle with the substrate. During the early stages of spreading and migration, cells can exert appreciable out-of-plane traction forces via rotation of focal adhesions~\cite{legant2013}. The quantity $T_i$ is a stress in three dimensions, i.e., a force per unit area. It describes the in-plane traction force per unit area that the cell exerts on the substrate. The force-balance equation is supplemented by the mass balance equation, such that cell density, $\rho({\bf r}_\perp,t)$, obeys the following conservation equation,
\begin{equation}
\partial_t \rho + \nabla.(\rho {\bf v})=\chi \rho\;,
\end{equation}
where ${\bf v}$ is the velocity field, and $\chi$ is the rate of variation in cell density due to cell division or death~\cite{bove2017}. In the following, we assume $\chi=0$. See refs~\cite{murray1984,ranft2010,yabunaka2017} for continuum models for tissues with explicit consideration of cell division and death.

\subsection{Constitutive model for intercellular stress}\label{subsec:1}
\begin{figure}[t]
\sidecaption
\includegraphics[scale=.65]{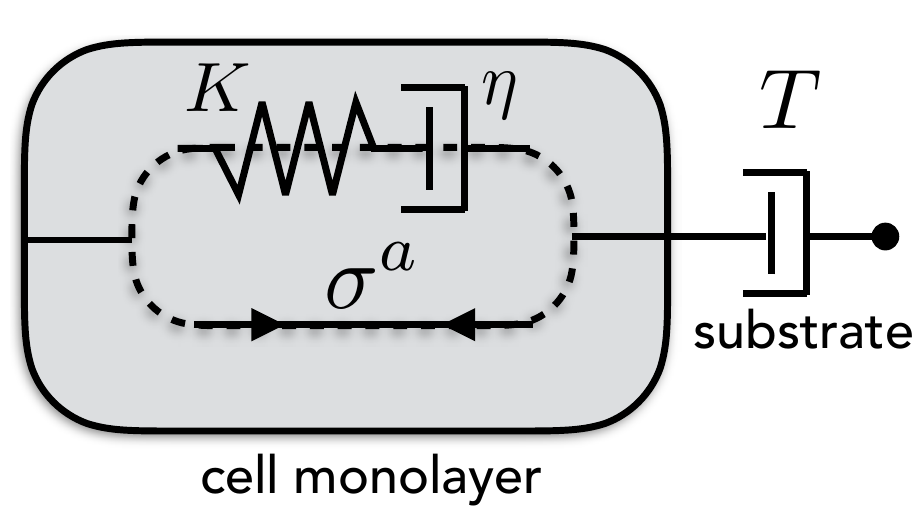}
\caption{{\bf Constitutive elements of the continuum model for collective migration}. The viscoelastic and active elements exert stresses in parallel. A local gradient in stress is balanced by the traction exerted by the cell on the substrate, which is modelled by a viscous element.}
\label{fig:2}       
\end{figure}

 The in-plane cellular stress, $\bm\sigma$, can be decomposed as the sum of intercellular stress, $\bm{\sigma^c}$, and active stress, $\bm{\sigma^{a}}$, originating from active intracellular processes (Fig.~\ref{fig:2}). 
The form of the constitutive relation for the intercellular stress has been highly debated, given the complex rheology of cellular aggregates~\cite{khalilgharibi2016}. On the timescale of seconds to minutes, living tissues behave elastically, recovering their original shape after a transient application of force~\cite{phillips1978,guevorkian2010}. On longer timescales (tens of minutes to hours), cellular aggregates exhibit fluid-like behavior that can arise from cell-cell adhesion turnover, cellular rearrangements, cell division or death ~\cite{ranft2010,guillot2013,heisenberg2013}. It is therefore commonly assumed that intercellular stresses obey Maxwell visco-elastic constitutive law~\cite{lee2011}, described by solid-like response at short time scales and fluid-like behavior at longer time scales. 

Experimental and computational studies by many groups have shown, however,  that stresses imposed on tissues cannot be completely dissipated, and cells support some part of applied tension~\cite{wayne2004,harris2012,gonzalez2013}. In fact rheological experiments have demonstrated that stress relaxation in epithelial monolayers can be described by a spring connected in parallel to a viscous dashpot~\cite{harris2012,khalilgharibi2018}. Others have shown that mechanical stress buildup in monolayers occurs in unison with strain accumulation~\cite{serra2012}, which can be described by an elastic constitutive law~\cite{tambe2011,mertz2012}. Therefore, to describe the dynamic mechanical behavior of cohesive cellular aggregates we assume linear Kelvin-Voigt rheology (Fig.~\ref{fig:2})~\cite{murray1984}
\begin{equation}\label{eq:stress}
\bm{\sigma^c}=(1+\tau \partial_t)\left[K \nabla.{\bf u}\ \mathbb{1}+ \mu\left(\nabla{\bf u} + (\nabla{\bf u})^T-\nabla.{\bf u}\ \mathbb{1}\right) \right]\;,
\end{equation}
where $\mathbb{1}$ is the identity matrix, ${\bf u}$ is the cellular displacement field, $K$ is the compressional elastic modulus, $\mu$ is the shear modulus, and $\tau$ is the viscoelastic relaxation timescale. The assumption of isotropic elasticity is consistent with stress measurement in cell monolayers using monolayer stress microscopy~\cite{notbohm2016,tambe2011}. For simplicity, we have ignored nonlinear contributions to the constitutive relation in Eq.~\eqref{eq:stress}, which may be essential for stabilizing the dynamical response of living tissues to large mechanical strain~\cite{banerjee2011c,kopf2013,banerjee2017}. In Sect.~\ref{sec:4}, we discuss the quantitative comparisons between elastic and fluid models of tissue rheology. We note that recent experimental studies show evidence for more complex rheological properties, including combinations of active elastic and dissipative response at moderate stretching~\cite{khalilgharibi2018}, as well as superelastic behavior at extreme stretching~\cite{latorre2018}.

\subsection{Active intracellular stress}\label{subsec:2}
The active intracellular stress stems from contractile forces generated in the actomyosin cytoskeleton in the cell cortex~\cite{murrell2015}, and from actin treadmiling driven by the assembly and diassembly actin filaments. Active contractile stresses depend on the concentration of actomyosin units, $c(t)$, with the form
\begin{equation}
\bm{\sigma^a}=\sigma_0(c) \mathbb{1} + \sigma_{an}(c){\bf p}{\bf p}\;,
\end{equation}
\begin{figure}[t]
\sidecaption[t]
\includegraphics[scale=.85]{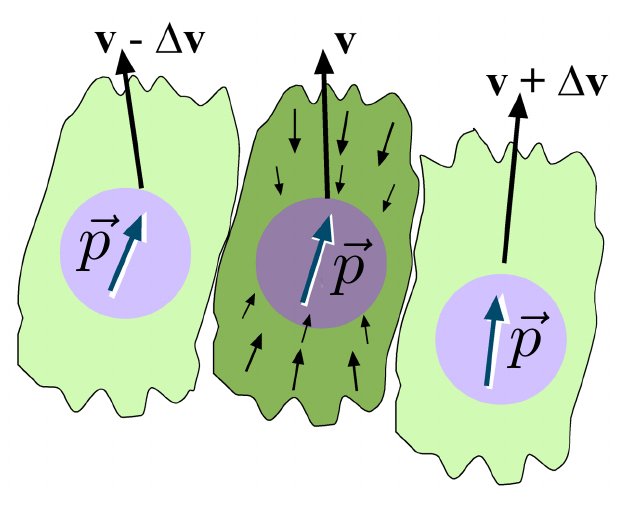}
\caption{{\bf Coordination of cell motion and polarization}. Cells align their motion along the polarity vector, ${\bf p}$, and move with a velocity ${\bf v}$. Neighboring cells tend to align their polarities, and poalrity differences generate a net torque on neighboring cells. Cells also exert a dipole-like contractile stress on the substrate due to actomyosin activity. Figure adapted from Ref.~\cite{lee2011}.}
\label{fig:3}       
\end{figure}
where we have introduced the cell polarization or polarity vector, ${\bf p}$, which is an internal state variable that controls the local direction of cell motion (Fig.~\ref{fig:3}). $\sigma_0(c)$ and $\sigma_{an}(c)$ are the isotropic and anisotropic components of the active stress due to actomyosin contractility. Note that additional active stress terms of the form $\propto \nabla {\bf p}$ are allowed by symmetry in this phenomenological model, leading to renormalization of the elastic modulus to leading order~\cite{banerjee2011}. Several models for the dependence of $\sigma_0$ on $c$ have been proposed, including linear~\cite{banerjee2011b}, logarithmic~\cite{banerjee2015} and saturating behaviour~\cite{bois2011}. Recent {\it in vitro} measurements show that contractile strains accumulate cooperatively as a function of myosin density~\cite{linsmeier2016}, indicating that $\sigma_0$ could take the general Hill functional form:
\begin{equation}
\sigma_0(c)=\sigma_0\frac{c^n}{c_*^n+c^n}\;,
\end{equation}  
where the constant $n>1$ indicates cooperative behavior beyond a critical concentration $c_*$, and $\sigma_0>0$ is the magnitude of the contractile stress. 

Finally, the force balance equation, Eq.~\ref{eq:force-balance}, requires a constitutive equation for the net traction stress transmitted to the substrate. For a layer of motile cells this is chosen of the form (Fig.~\ref{fig:1}c)~\cite{banerjee2015}
\begin{equation}
{\bf T}=\zeta {\bf v} - f{\bf p}\;,
\label{eq:traction}
\end{equation}
where ${\bf v}=\partial_t{\bf u}$, $f$ is the magnitude of the propulsion force, and $\zeta$ is an effective friction coefficient that depends on the rate of focal adhesion turnover~\cite{walcott2010}. This form for traction in Eq.~\eqref{eq:traction} results in local misalignment of traction stress and cell velocity, consistent with experimental findings~\cite{notbohm2016,brugues2014}. The propulsion force, $f{\bf p}$, drives cell crawling, and depends on the concentration of branched actin in the lamellipodia of migrating cells.  For simplicity, we assume that there is a steady concentration of polymerized actin that pushes the cell forward. Dynamic models for the competition between branched and contractile actin have been proposed~\cite{lomakin2015,suarez2016}. A detailed description of such molecular processes lies beyond the scope of this review, but can be easily incorporated within this framework. The resultant force balance equation is then given by (Fig.~\ref{fig:1}c,\ref{fig:2}),
\begin{equation}\label{eq:s}
h\nabla.(\bm{\sigma^c}+\bm{\sigma^a})=\zeta {\bf v} - f{\bf p} + {\bf f}_\text{ext}\;,
\end{equation}
where ${\bf f}_\text{ext}$ is the external force (density) applied to the system. In the absence of external forces or stresses applied at the boundary, the net traction force, when integrated over the entire cell-substrate interface must vanish. This implies a fundamental constraint on the relatioship between cell polarity and velocity: 
\begin{equation}
\int {\bf v}.{\bf dA}=\frac{f}{\zeta}\int {\bf p}.{\bf dA}\;.
\end{equation}
In the following, we will additionally need to prescribe the dynamics of cell polarization and actomyosin concentration, which regulate active cell motility and the production of contractile stresses.

\subsection{Mechanochemical coupling of cell motion and contractility}\label{subsec:3}
The dynamics of cell polarization is commonly modeled following the physics of active liquid crystals~\cite{marchetti2013}, a phenomenological approach that requires further justification and scrutiny. The cell polarization vector evolves in time according to,
\begin{equation}\label{eq:p}
\partial_t {\bf p} + \beta ({\bf p}.\nabla){\bf p} + {\bf v}.\nabla {\bf v} - \frac{1}{2}(\nabla \times {\bf v})\times {\bf v}=a(1-\vert {\bf p}\vert^2){\bf p} + \kappa \nabla^2 {\bf p} + w \nabla c\;,
\end{equation}
where the advective coupling $\beta$ arises from ATP driven processes such as treadmiling~\cite{ahmadi2006}, the velocity dependent advective terms are borrowed from the nematic liquid crystal literature~\cite{prost1995}, and the Franck elastic constants are both assumed to be equal to $\kappa$. Here, $a$ controls the rate of relaxation to a homogeneously polarized cell monolayer, and $\kappa$ controls the strength of nearest-neighbor alignment of the polarization field (Fig.~\ref{fig:3}), akin to velocity alignment in the Viscek model of collective motion~\cite{vicsek1995}. The active mechanochemical coupling $w > 0$ represents the rate of alignment of cell polarization with gradients in the actomyosin concentration field. As a result, local cell motion is guided toward regions of high contractility.

\begin{figure}[t]
\sidecaption
\includegraphics[scale=.4]{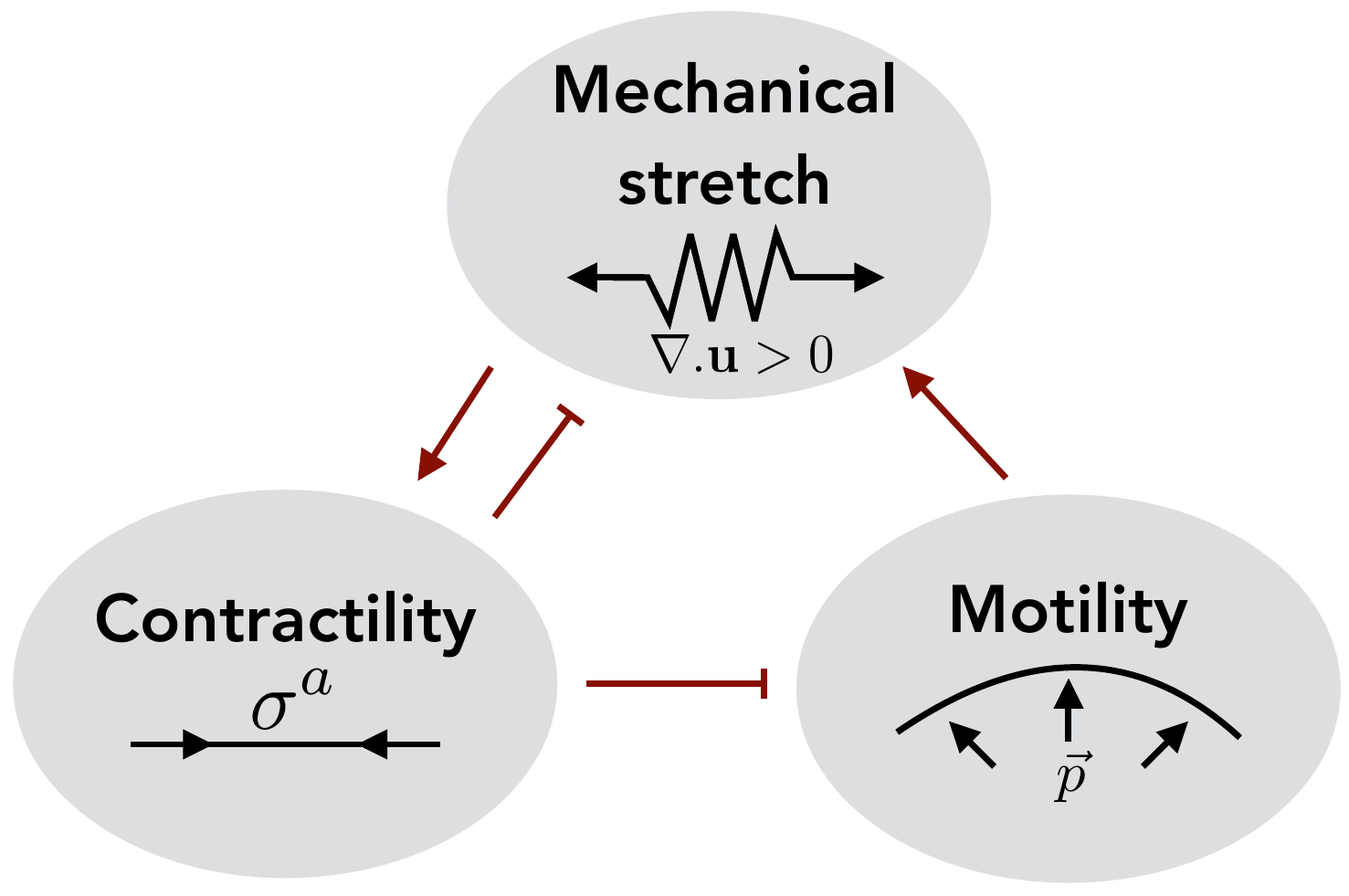}
\caption{{\bf Mechanochemical feedback mechanisms}. Feedback between cell stretch, actomyosin contractility and polarized cell motility in the mechanochemical model for collective motion. Local stretch upregulates assembly of actomyosin, which generates contractile forces that exert compressive stress. Polarized motility, in turn, pulls and stretches the cells.}
\label{fig:4}       
\end{figure}

The concentration of contractile actomyosin is described by a reaction-advection-diffusion equation,
\begin{equation}\label{eq:c}
\partial_t c + \nabla. (c{\bf v}) = D \nabla^2 c -\frac{1}{\tau_c}(c-c_0) + \alpha c_0 \frac{\nabla.{\bf u}}{1+\vert \nabla.{\bf u}\vert/s_0}\;,
\end{equation}
where $D$ is a diffusion constant, $\tau_c$ is the timescale of relaxation to steady-state, and $\alpha > 0$ is the rate of accumulation of contractile actomyosin due to local tissue stretching~\cite{banerjee2015}. The positive constant $s_0$ sets the upper limit of strain magnitude above which the production rate of $c$ saturates~\cite{kopf2013}. This mechanochemical feedback (Fig.~\ref{fig:4}) is consistent with experimental data for single cells~\cite{robin2016} and cell monolayers~\cite{serra2012, vincent2015}, where a local extensile strain reinforces contractility via assembly of actomyosin~\cite{levayer2012}. Turnover of contractile elements at a rate $\tau_c^{-1}$ fluidizes the monolayer, inducing an effective viscosity of magnitude $\eta_\text{eff}=(K-\sigma_0 + D\zeta/h)\tau_c$~\cite{banerjee2015}. Aside from the negative feedback between mechanical strain and actomyosin assembly, positive feedback occurs between mechanical strain and advective fluxes into regions of high contractility. Advective transport can compete with diffusion to generate steady state patterns of contractility~\cite{gross2017}.

It is instructive to note that for small changes in $c$ around $c_0$, Eq.~\eqref{eq:c} describes a dynamics of active contractile stress that is similar to a Maxwell constitutive model for intercellular stress proposed by Lee and Wolgemuth~\cite{lee2011}. Here, in addition, we consider an elastic contribution to the active stress, described by the term $\alpha$. The feedback between mechanical strain and contractility yields an effective elastic modulus $K_\text{eff}\approx K+\alpha\tau_c (\sigma_0+fw/2ah)$~\cite{banerjee2015}, larger than the modulus $K$ of the monolayer in the absence of contractility. This prediction is consistent with experimental measurements that cell monolayers  treated with blebbistatin (myosin-II inhibitor) have a much reduced elastic modulus~\cite{notbohm2016}.

\section{Forces and motion driving collective cell behavior}\label{sec:3}
The coupled system of Eqs.~\eqref{eq:s}-\eqref{eq:c} describes the spatiotemporal dynamics of cell monolayers, subject to appropriate boundary and initial conditions for cellular displacement (${\bf u}$), cell polarity field (${\bf p}$) and  actomyosin concentration ($c$). We now discuss the quantitative predictions of this model for collective mechanics and migration  in various biological contexts. In particular we will focus on four scenarios where continuum model predictions have been tested and validated against experimental data: {\it Force transmission in epithelial monolayers} (Sect.~\ref{subsec:3-1}), {\it Collective motility in expanding monolayers} (Sect.~\ref{subsec:3-2}), {\it Cell migration under confinement} (Sect.~\ref{subsec:3-3}), and {\it Epithelial movement during gap closure} (Sect.~\ref{subsec:3-4}). 

\subsection{Force transmission in epithelial monolayers}\label{subsec:3-1}
Epithelial cell monolayers adherent to soft elastic substrates provide a model system for mechanical force generation during tissue growth, migration and wound healing~\cite{wozniak2009,ladoux2017}. In the experimental assays of interest~\cite{du2005,trepat2009}, the substrates are usually coated with extracellular matrix proteins (e.g. fibronectin, collagen) that allow cells to spread fully to a thin film and thereby establish contractile tension. To describe the experimentally observed traction force localization in fully spread adherent cell sheets~\cite{du2005,trepat2009,mertz2012}, we consider the steady-state limit of Eq.~\eqref{eq:s}-\eqref{eq:c}, which was originally studied in refs~\cite{edwards2011,banerjee2011,banerjee2012,mertz2012}. In this limit, ${\bf v}\equiv 0$, and concentration of active contractile units is slaved to material strain, $c\approx c_0(1+ \alpha\tau_c\nabla.{\bf u})$. This results in renormalization of the compressional modulus to linear order. Similarly from Eq.~\eqref{eq:p} it follows that ${\bf p} \approx - \left(\frac{w\alpha \tau_cc_0}{\kappa}\right){\bf u}$. 

To linear order, the force balance equation for the contracting cell layer, with internal stress $\bm{\sigma}=\bm{\sigma^c}+\sigma_0 \mathbb{1}$, is given by,
\begin{equation}\label{eq:s1}
h\nabla.\bm{\sigma}=Y{\bf u}\;,
\end{equation}
where, $Y=k+\frac{fw\alpha \tau_cc_0}{\kappa}$ is the effective substrate rigidity, resulting from the sum of substrate stiffness $k$, and the contribution from cell polarization. The intercellular stress, $\bm{\sigma^c}$, follows a constitutive relation identical to that of a linear elastic solid with a renormalized compressional modulus $K_\text{eff}$. Equation~\eqref{eq:s1} can be exactly solved for circularly shaped monolayers~\cite{edwards2011,mertz2012}, subject to the stress-free boundary condition: ${\bm \sigma}.\hat{{\bf n}}=0$, where $\hat{{\bf n}}$ is the unit normal to the boundary of the monolayer. This boundary condition needs to be appropriately modified if the colony edge is under tension due to peripheral actin structures~\cite{ravasio2015}.

\begin{figure}[b]
\sidecaption
\includegraphics[width=\columnwidth]{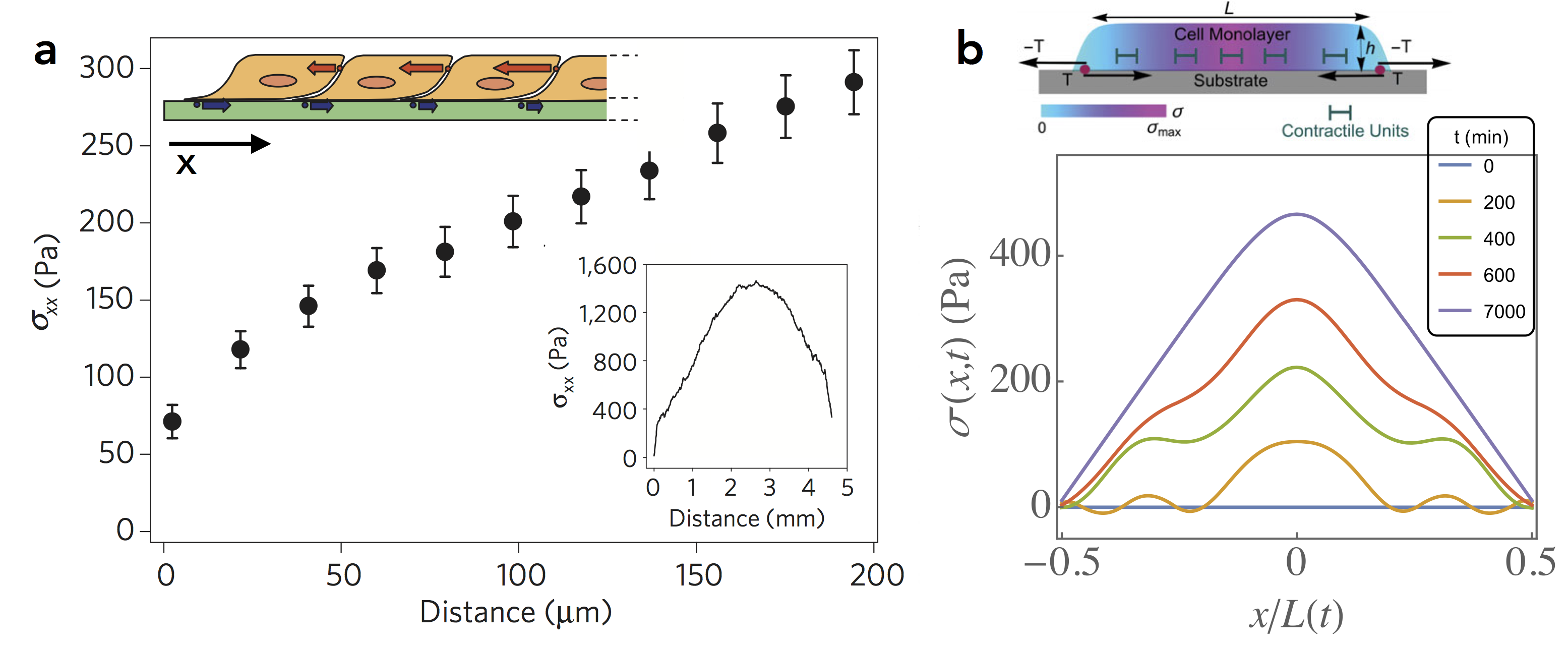}
\caption{{\bf Stress transmission in epithelial monolayers.} (a) Internal stress, $\sigma_{xx}$, in an expanding MDCK monolayer obtained by integrating cellular traction force. Adapted from Ref.~\cite{trepat2009}. Buildup of $\sigma_{xx}$ signifies that tension in the actin cytoskeleton and cell-cell junctions increases towards the centre of the monolayer. (b) Time evolution of the internal stress $\sigma(x,t)$ in the monolayer predicted by the continuum model of epithelium~\cite{banerjee2015}.}
\label{fig:5}       
\end{figure}
The resulting solution to Eq.~\eqref{eq:s1} describes cell traction forces and displacements localized to the edge of the monolayer over a a length scale $\ell_p=\sqrt{K_\text{eff}h/Y}$, defined as the {\it stress penetration depth}. Furthermore, internal stresses in the monolayer, $\bm{\sigma}$, accumulate at the center of the monolayer, in agreement with experimental data (Fig.~\ref{fig:5}a-b)~\cite{trepat2009,tambe2011}. The model can be solved numerically for monolayers of any geometry, and it predicts that traction stresses localize to regions of high curvature~\cite{banerjee2013}. This was later confirmed experimentally by micropatterning adhesion geometries of non-uniform curvatures~\cite{oakes2014}. The model has been used to recapitulate a number of experimental observations~\cite{banerjee2011,banerjee2012,banerjee2013,mertz2012}, including substrate rigidity dependence of traction stresses~\cite{ghibaudo2008} and cell spread area~\cite{chopra2011}, traction stress dependence on cell geometry~\cite{oakes2014}, correlation between cell shape and mechanical stress anisotropy~\cite{roca2008}, as well as the optimal substrate rigidity for maximal cell polarization~\cite{zemel2010}.

\begin{figure}[t]
\centering
\sidecaption
\includegraphics[scale=.65]{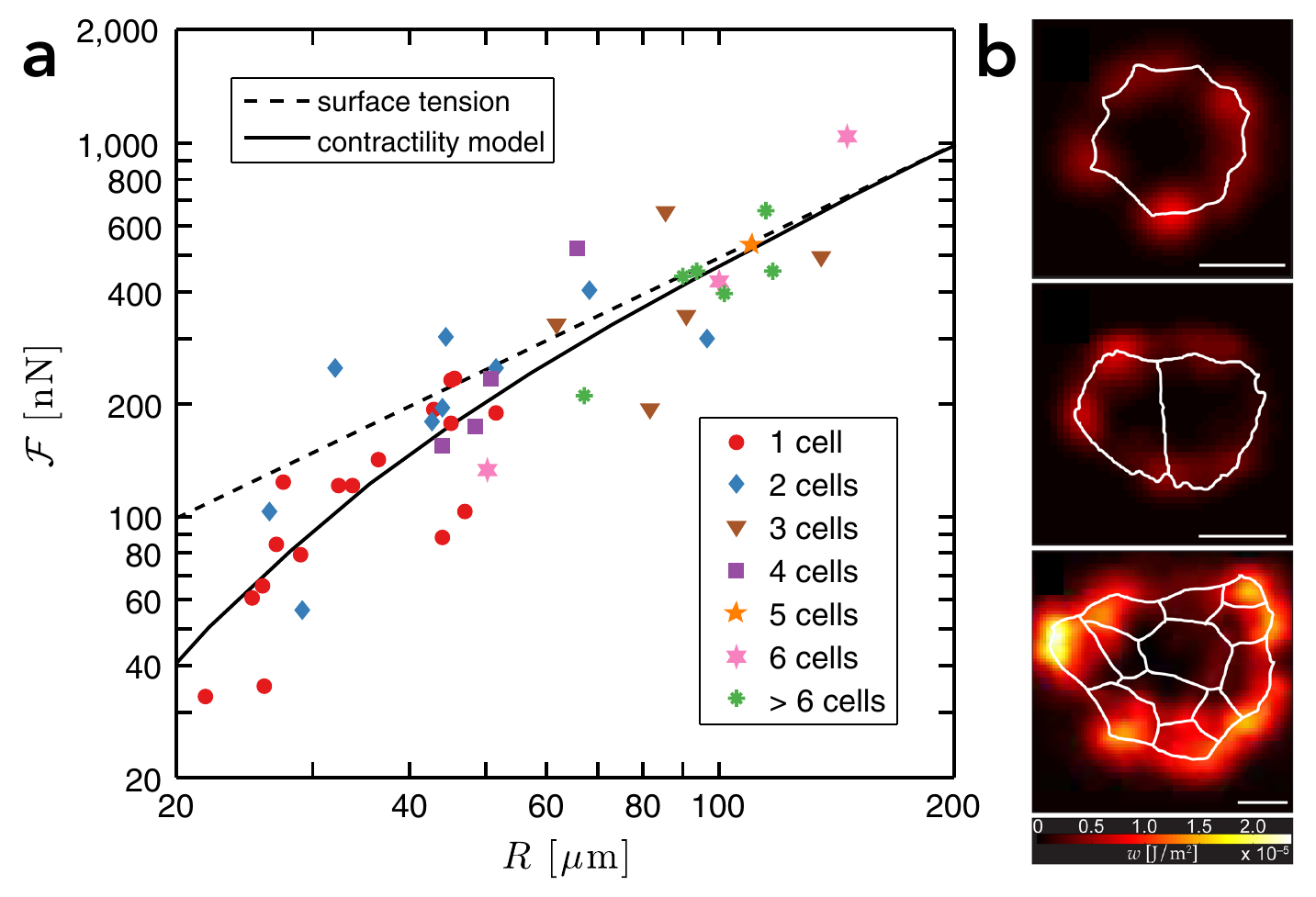}
\caption{{\bf Active surface tension in cohesive epithelial colonies}. (a) Total force transmitted to the substrate by keratinocyte colonies, $\mathcal{F}$, as a function of the equivalent radius, $R$, of the colonies~\cite{mertz2012}. The dashed line represents the linear scaling expected for surface tension, $\mathcal{F}\propto R$. The solid line shows a fit of the data to the continuum model in Eq.~\eqref{eq:s1}. (b) Distribution of strain energy, $w$, for a representative single cell, pair of cells, and colony of 12 cells. Scale bar=50 $\mu$m.}
\label{fig:6}       
\end{figure}
A particularly interesting application of this model is in understanding the relationship between traction force magnitude and the geometric size of cohesive cell colonies adherent to soft matrices~\cite{mertz2012}. One can define the magnitude of the total traction force transmitted to substrate as $\mathcal{F}=\int \vert {\bf T}.{\bf dA}\vert$, where the integral is taken over the entire spread area of the colony, $A$. The model predicts that for large cell colonies of linear size $R\gg \ell_p$, $\mathcal{F}=2\pi h\sigma_0 R$. This linear scaling of force with colony size (Fig.~\ref{fig:6}) implies that actomyosin contractility, $\sigma_0$, induces an effective surface tension in solid tissues, which appear to wet the substrate underneath akin to fluid droplets. The effective surface tension was estimated from experiments on keratinocyte colonies to be $8\times 10^{-4}$ N/m~\cite{mertz2012}, which is of the same order of magnitude as the apparent surface tension estimated in adherent endothelial cells~\cite{bischofs2009}, Dictyostelium cells~\cite{delanoe2010}, mm-scale migrating epithelial sheets~\cite{trepat2009}, and cellularised aggregates~\cite{guevorkian2010}. Recent work has shown that for highly motile and fluid cell colonies, traction forces localize to the colony interior rather than at the edge~\cite{schaumann2018}.

\subsection{Collective motility in expanding monolayers}\label{subsec:3-2}
Migratory behaviors of epithelial cells are commonly studied experimentally using the wound healing assay. In the classical {\it scratch-assay}~\cite{yarrow2004}, a strip of cells is removed from the monolayer to observe collective migration of cells marching to fill the tissue gap. This experimental model system, however, is unsuited for controlled study of migration due to ill-defined borders and debris created by the physical wound. The last decade has seen significant improvement in the wound healing assay, where cells are grown to confluence within a removable barrier, which is then lifted to allow cell migration into free space~\cite{poujade2007,trepat2009}. These studies, in combination with Traction Force Microscopy have shed light into the forces and motion driving collective cell migration. In particular, it has been observed that cell velocity fields at the leading edge of the epithelium exhibit complex swirling patterns~\cite{Petitjean2010} and often form multicellular {\it migration fingers}~\cite{poujade2007}. Measurement of mechanical stresses at cell-cell and cell-substrate interfaces have given rise to models of {\it tug-of-war}~\cite{trepat2009}, a consequence of mechanical force-balance, where local traction stresses in the monolayers are integrated into long-ranged gradients of intercellular tensions (Fig.~\ref{fig:5}a-b). Stress inference at cell-cell junctions have led to the suggestion of {\it plithotaxis}~\cite{tambe2011}, where cell migration is guided towards the direction of maximum normal stress and minimum shear stress.
\begin{figure}[t]
\sidecaption[t]
\includegraphics[width=\columnwidth]{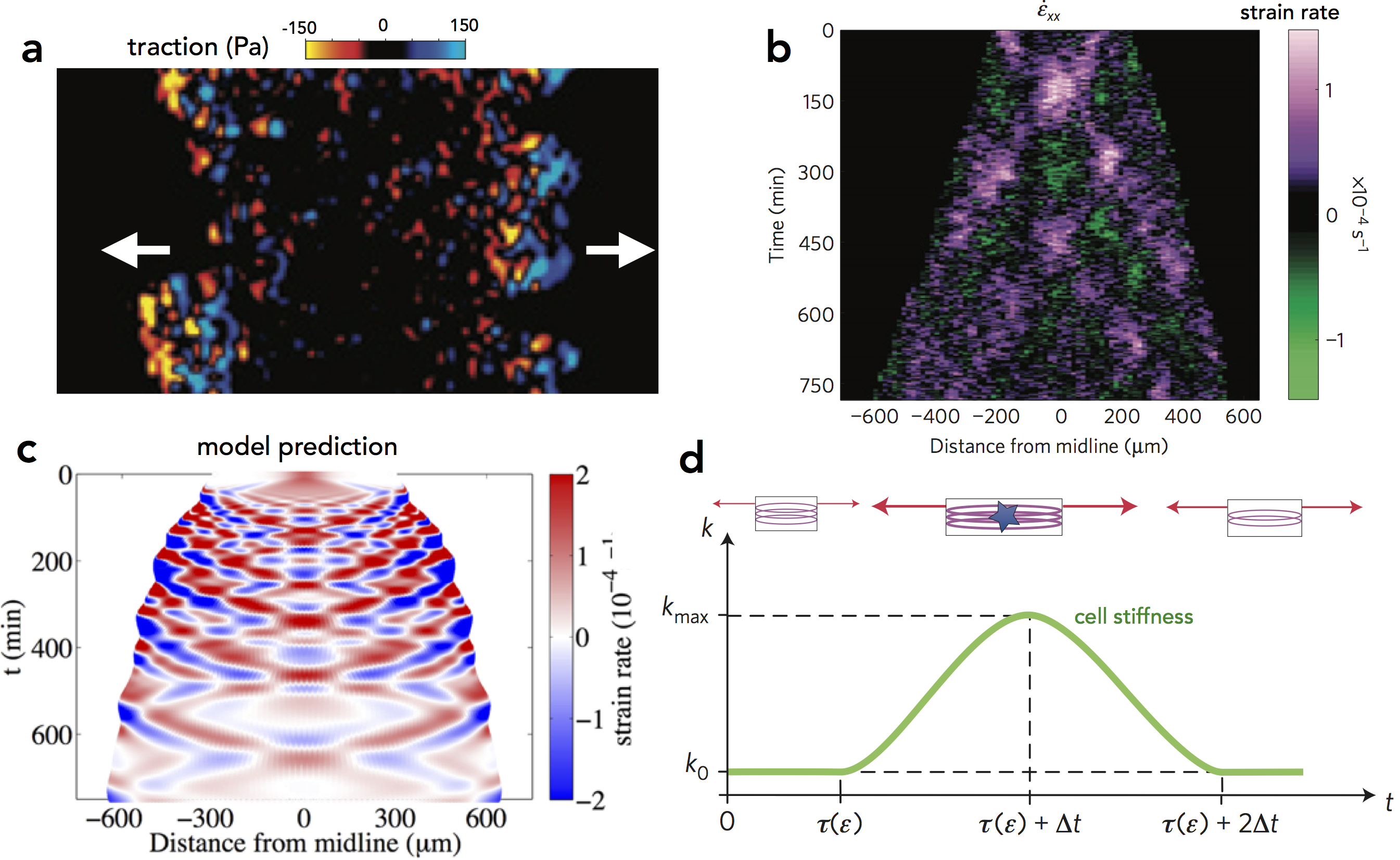}
\caption{{\bf Mechanical waves during epithelial expansion.} (a) Traction stress map of an expanding MDCK monolayer, adapted from Ref.~\cite{serra2012}. (b) Kymograph of strain rate in expanding MDCK monolayers~\cite{serra2012}, showing generation and propagation of X-shaped mechanical waves. (c) Propagating stress waves predicted by the continuum model, Eq.~\eqref{eq:s}-\eqref{eq:c}~\cite{banerjee2015}. (d) Schematic illustrating the mechanics of migration waves, adapted from refs.~\cite{thery2012,serra2012}. Cells at the colony center (purple) are initially stretched by pulling forces generated by leader cells. Stretched cells recover their equilibrium shape via cytoskeletal fluidization (blue star), which is then reinforced to trigger shape elongation again. These shape oscillations mediate periodic stiffening and fluidization of cells (green curve).}
\label{fig:7}       
\end{figure}

A particularly interesting case is that of collective migration waves, observed in mm-sized monolayers expanding into free space~\cite{serra2012} (Fig.~\ref{fig:7}a). These mechanical waves, crucially dependent on myosin contractility and cell-cell adhesions, propagate at a slow speed (on the order of $\mu$m/hr) from the colony edge to the center and back (Fig.~\ref{fig:7}b). The waves are mediated by shape changes at the scale of single cells. Pulling forces from crawling cells at the leading edge of the colony stretch interior cells, which periodically recover their shape via a proposed model of cytoskeletal fluidization (Fig.~\ref{fig:7}d)~\cite{thery2012}. Interestingly, this wave-like progression of cell movement naturally arises in the active elastic media models, Eq.~\eqref{eq:s}-\eqref{eq:c}, due to a feedback between contractility and mechanical strain~\cite{banerjee2015}. 

To understand the origin of wave propagation and estimate the wave frequency, it is useful to examine the mechanics of an expanding one-dimensional monolayer with a polarization field pointing outward from the colony center. We consider the linear fluctuations in the strain field $\delta \varepsilon$ and the concentration field $\delta c$, about the quiescent homogeneous state $u=0$, $c=c_0$. Using Eqs. \eqref{eq:s} and \eqref{eq:c}, one can eliminate $\delta c$ to obtain the linearized dynamics of strain fluctuations:
\begin{equation}
\tau_c \zeta\partial_t^2 \delta\varepsilon + \zeta \partial_t\delta \varepsilon=h(K_\text{eff} + \eta_\text{eff}\partial_t -\tau_c KD\partial_x^2)\partial_x^2\delta\varepsilon\;.
\end{equation}
The above equation shows that the coupling of strain to concentration field yields an effective mass density (inertia), $\tau_c\zeta$, and viscoelasticity characterized by an effective elastic modulus, $K_\text{eff}$, and an effective viscosity $\eta_\text{eff}$, which leads to oscillations with a characteristic frequency $\omega=q\sqrt{h(K_\text{eff}+\tau_cq^2KD)/(\tau_c\zeta)}$, with $q$ the wavevector. Full solutions of the nonlinear equations~\cite{banerjee2015} yields X-shaped propagating stress waves akin to experimental data (Fig.~\ref{fig:7}c)~\cite{serra2012}. These contraction waves are chraracterized by sustained oscillations in tissue rigidity - a slow period of stiffening followed by rapid fluidization (Fig.~\ref{fig:7}d). When the coupling of polarization to strain and contractility is turned on, complex spatiotemporal patterns emerge including traveling stress pulses and chaotic polarization waves~\cite{kopf2013,banerjee2015}.

\subsection{Cell migration under confinement}\label{subsec:3-3}

\begin{figure}[t]
\centering
\sidecaption
\includegraphics[width=0.85\columnwidth]{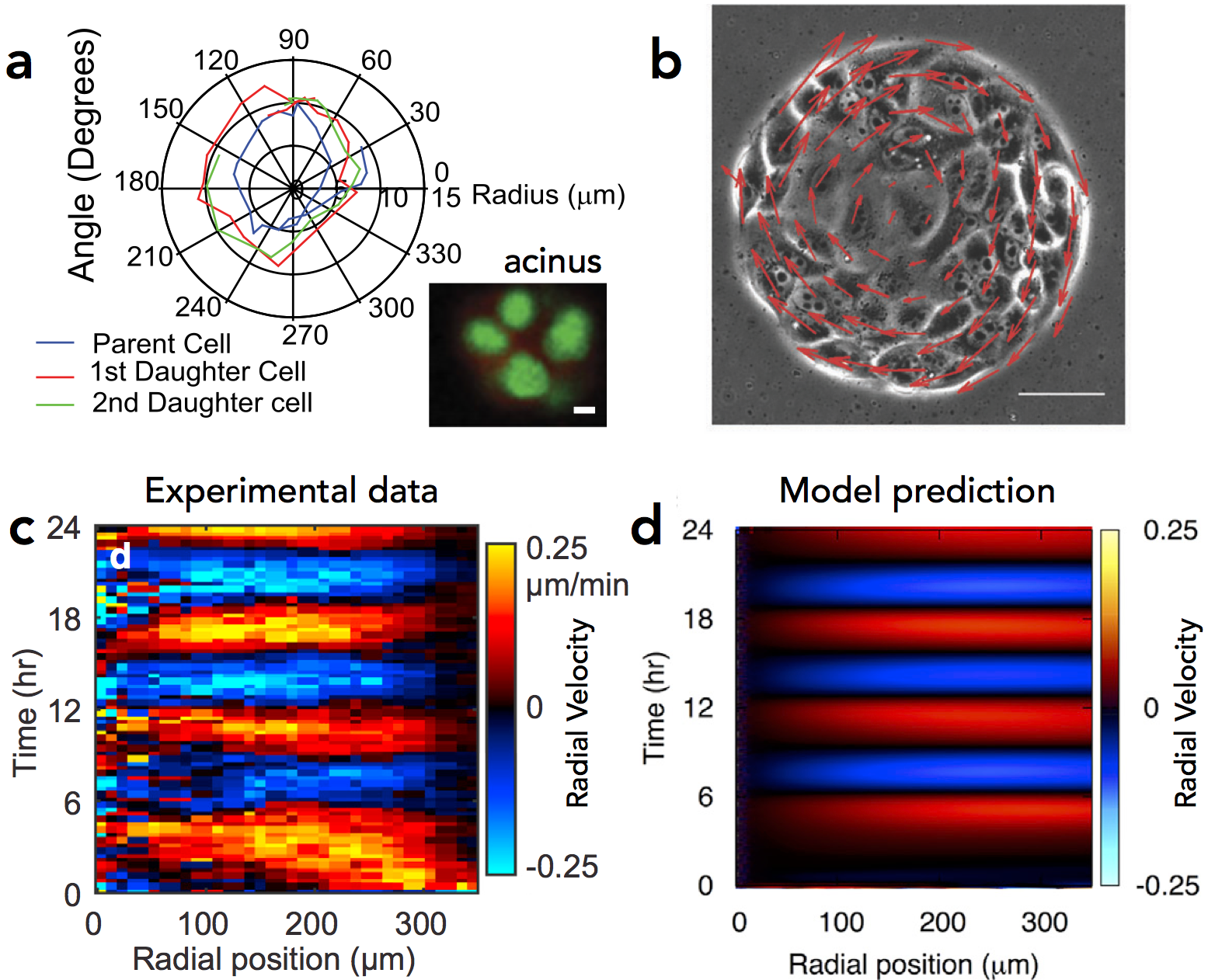}
\caption{{\bf Coherent cell motion in confined environment}. (a) {\it Coherent angular motion} of cells during acinus morphogenesis, adapted from Ref.~\cite{tanner2012}. Graph shows angular rotation of the parent and daughter cells obtained by nuclei tracking. Inset: Cross section of acinus with F-actin staining in green (Scale bar=30 $\mu$m). (b) Collective rotation of MDCK cells seeded on circular fibronectin patterns, reproduced from Ref.~\cite{doxzen2013}. The magnitude and the direction of local velocity fields are indicated by red arrows (Scale bar = 50 $\mu$m). (c) Kymograph of radial velocity fields of confluent cells in a micropattern~\cite{notbohm2016}, showing periodic oscillations. (d) Radial velocity kymograph, obtained by simulating Eqs.~\eqref{eq:s}-\eqref{eq:c}, reproducing collective cell oscillations.}
\label{fig:8}       
\end{figure}

In many biological contexts, including morphogenesis, tissue polarity establishment, and acini formation, cells often migrate collectively in confined environments. Experiments have shown the emergence of coherent angular motion of cells {\it in vivo} (Fig.~\ref{fig:8}a), including cells in yolk syncytial layer of zebrafish embryos~\cite{d2001}, and breast epithelial cells in 3D collagen gels~\cite{tanner2012}. These self-generated persistent motions are crucially dependent on cell-cell adhesions and myosin contractility, loss of which can drive malignant behavior. In recent years, collective motion in geometric confinement have been studied in a more controlled manner using adhesive micropatterns~\cite{thery2009}, which allow confinement of cell cultures in geometric domains of any shape and size. 

When plated in circular micropatterns, small sized epithelial monolayers often exhibit large scale correlated movements and spontaneous swirling motions, as shown in Fig.~\ref{fig:8}b~\cite{doxzen2013,deforet2014,segerer2015,notbohm2016}. These collective rotations emerge once the cells have reached a critical density (2000 cells/mm$^2$) and occur in micropatterns of radii smaller than the cellular velocity correlation length ($\sim 200$ $\mu$m) in unconfined situations~\cite{doxzen2013}. Furthermore these rotations require cell-cell adhesions for efficient transmission of motility cues by contact guidance~\cite{doxzen2013}, and radial velocity oscillations are observed with a time period linearly proportional to the micropattern radius~\cite{deforet2014}. Aside from collective rotational motion, emergence of active nematic states has also been observed in confined monolayers of elongated fibroblasts and MDCK cells~\cite{duclos2014,duclos2017,saw2017}. In these cases, cells actively transfer alignment cues from the boundary to the bulk of the monolayer, resulting in domains of alignment and topological defect patterns.

Different cell-based computational models have been implemented to recapitulate collective rotational motion, including the the cellular Potts model~\cite{kabla2012,doxzen2013,albert2016}, active particle models~\cite{deforet2014}, and voronoi-type models~\cite{li2014}, where persistent rotations emerge due to velocity alignment mechanisms of motile cells. In recent work~\cite{notbohm2016}, we described collective rotations using a continuum model similar to Eq.~\eqref{eq:s}-\eqref{eq:c} (Fig.~\ref{fig:8}c-d). This model quantitatively captures a key aspect of the experimental data, namely, that the cell velocity field alternated between inward and outward radial motion with a time period equal to that of the oscillations in the intercellular stress~\cite{notbohm2016}. This wave-like motion is predicted by the model to arise through the chemomechanical feedback between the mechanical strain, $\nabla{\bf u}$, and actomyosin contractility, $c$~\cite{banerjee2015}. In the limiting case where cell deformations, ${\bf u}$, is only coupled to polarity ${\bf p}$, no oscillatory behavior is observed. This prediction was confirmed by experiments, where inhibition of contractility by blebbistatin eliminated the multicellular oscillations. Furthermore, the polarization field, {\bf p}, is crucial to capture the misalignment between traction and velocity, observed experimentally. Overall, the coupling of cell motion to polarization and actomyosin contractility is required to to capture the experimentally observed distribution of traction forces~\cite{notbohm2016}, which points inward at the periphery of the micropattern and oscillates between outward and inward within the bulk of the monolayer. 

\subsection{Epithelial movement during gap closure}\label{subsec:3-4}

\begin{figure}[t]
\centering
\sidecaption
\includegraphics[width=0.85\columnwidth]{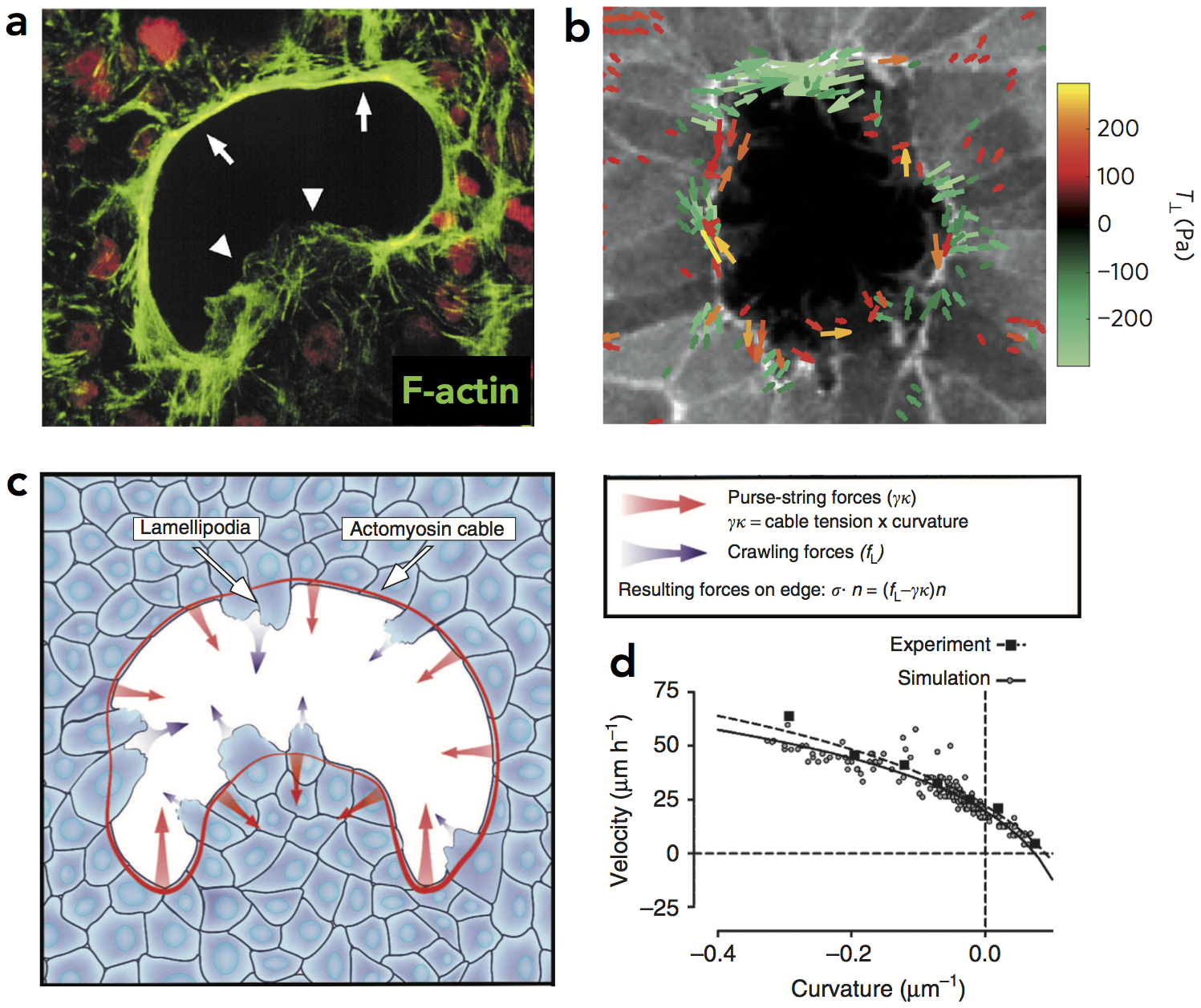}
\caption{{\bf Collective migration during epithelial gap closure}. (a) Closure of {\it in vitro} wound in epithelial monolayers is mediated by a combination of purse-string based contraction of actomyosin cable (arrows) and lamellipodia based cell crawling (arrowheads). Figure adapted from Ref.~\cite{jacinto2001}. (b) Lamellipodial protrusions generate traction forces away from the wound (red arrows), whereas traction generated by purse-string based contraction point towards the wound (green arrows). Traction stress map reproduced from Ref.~\cite{brugues2014}. (c) Schematic of a continuum model for gap closure, showing the dependence of purse-string and crawling forces on the local gap geometry. (d) Migration velocity increases with increasing magnitude of local gap curvature. Reproduced from Ref.~\cite{ravasio2015b}.}
\label{fig:9}       
\end{figure}

Collective cell movement during epithelial gap closure is essential for maintaining the tissue mechanical integrity and to protect the internal environment from the outside by regenerating a physical barrier. Gaps can occur autonomously during development \cite{wood2002}, or can be generated by cell apoptosis \cite{rosenblatt2001} or tissue injury. It is widely accepted that epithelial gap closure is driven by two distinct mechanisms for collective cell movement (Fig.~\ref{fig:9}a)~\cite{jacinto2001,begnaud2016}. First, cells both proximal and distal to the gap can crawl by lamellipodial protrusions~\cite{martin1992,anon2012,fenteany2000}. Secondly, cells around the gap can assemble a multicellular actomyosin purse-string, which closes gaps via contractile forces (Fig.~\ref{fig:9}b) \cite{martin1992,bement1993}. The continuum framework described in this review can be appropriately adapted to study the relative contributions of crawling and contractile forces on epithelial gap closure.

Continuum models of tissue gap closure have considered both visco-elastic solid~\cite{vedula2015} and fluid~\cite{cochet2014,ravasio2015b} models of tissues. In either scenarios, force balance between cell-cell and cell-substrate interactions can be expressed as,
\begin{equation}\label{eq:str}
h\nabla.\bm{\sigma}=\zeta {\bf v}-f{\bf p}\;,
\end{equation}
where $f$ is the magnitude of the propulsion force acting on the cells due to lamellipodial protrusions, both proximal and distal to the gap, such that ${\bf p}$ points into free space. While previous continuum models have neglected the polarity term in the force balance, this is necessary for the misalignment of traction force and velocity observed for instance in closed contour wound healing assays~\cite{brugues2014}. To model the active pulling forces on the gap boundary, Eq.~\eqref{eq:str} is solved subject to the following boundary condition for the stress tensor on the moving gap boundary (Fig.~\ref{fig:9}c):
\begin{equation}
\bm{\sigma}.{\bf \hat{n}}=(f_L-\lambda \kappa){\bf \hat{n}}\;,
\end{equation}
where ${\bf \hat{n}}$ is the local normal vector on the gap boundary, directed away from the tissue, $f_L$ is the force density due to lamellipodial protrusions, $\kappa$ is the local gap boundary curvature (negative for circular gaps), and $\lambda$ is the line tension due to actomyosin purse-string. The model has been used to capture the sensitivity of collective motion on the local gap geometry~\cite{ravasio2015b} (Fig.~\ref{fig:9}d). For instance, crawling mediated migration ($\lambda=0$) occurs at a speed independent of gap curvature, whereas purely purse-string driven motility ($f_L=0$) increases with decreasing radius of curvature. This may explain why purse-string is not assembled for large wounds, as its driving force is inversely proportional to the gap diameter. A model of cable reinforcement, where tension $\lambda \propto \kappa$, has also been proposed to account for the experimentally observed increase in closure velocity and traction stress with time~\cite{vedula2015}. A more comprehensive model of gap closure dynamics with spatiotemporal variations in lamellipodia and purse-string forces (Fig.~\ref{fig:9}b) has recently been implemented using the vertex model~\cite{staddon2018}. 

\section{Comparisons between active elastic and fluid models of collective cell migration}\label{sec:4}

Previous work has employed both elastic~\cite{kopf2013,banerjee2015} and fluid~\cite{arciero2011,lee2011,recho2016,blanch2017,blanch2017-2,perez2018} models of epithelial cell sheet to describe the dynamics of epithelial expansion, as probed for instance in wound healing assays (Fig.~\ref{fig:7}a). Both models can account for traveling waves, as observed in experiments, provided the sheet rheology is coupled to internal dynamical degrees of freedom, such as contractile activity (elastic model ~\cite{banerjee2015}) or cell division or polarization (fluid model~\cite{recho2016,blanch2017-2}). On the other hand, tissues can undergo fluidization/stiffening cycles, respond elastically or viscously on different times scales, and there is still no  continuum model capable of capturing their rheology across all time scales. 

In this section we compare the viscous and elastic continuum approaches for modeling cell monolayers, focusing on a one-dimensional ($1d$) model that allows for an analytical solution. The $1d$ calculation can also be directly compared to  experiments such as those shown in Fig.~\ref{fig:5}a, where the monolayer properties are generally averaged over the direction transverse to that of mean motion. Denoting by $x$ the direction of monolayer expansion, the in-plane force balance equation is simply given by
\begin{equation}
\zeta v_x=fp_x+h\partial_x\sigma\;,
\label{eq:balance_1d}
\end{equation}
where $\sigma=\sigma_{xx}=\sigma^c+\sigma^a$. In the absence of cell division and tissue growth, the volume of the monolayer remains approximately constant during expansion. This requires the product $L(t)h(t)$ to remain constant, where $L(t)$ and $h(t)$ are the monolayer width in the direction of expansion and the monolayer thickness at time $t$, respectively. For simplicity in the following we neglect the variation of $h$. 

To illustrate the difference between the fluid and elastic models we examine below the accumulation of contractile stresses in an isotropic expanding monolayer, with vanishing net polarization, that was discussed for the fluid case in Ref.~\cite{blanch2017}. In contrast to Ref.~\cite{blanch2017} we assume $\sigma^a=\text{constant}$, to incorporate contractile cell activity. We neglect both nonlinear active stresses and spatiotemporal variations of the concentration $c$ of contractile actomyosin. We additionally use a quasi-static approximation for the cell polarization that is assumed to relax on time scales much faster than those associated with cellular deformations and rearrangements. Finally, for simplicity we will neglect the spatial and temporal variation of the thickness $h$ of the monolayer. 
We retain only linear terms so that the polarization field, $p_x$,  satisfies the equation~\cite{blanch2017}
\begin{equation}
L_p^2\partial_x^2p_x=p_x\;,
\label{eq:pol_1d}
\end{equation}
where we have introduced the length scale $L_p=\sqrt{\kappa/a}$ that describes spatial variation in polarization within the monolayer.

The viscous or elastic nature of the cell sheet will be specified by the chosen form of the constitutive equation for the intercellular stresses, $\sigma^c$. One important distinction, not apparent in the linear form of the equations considered here, is that the fluid motion is treated in an {\it Eulerian} frame, while the elastic medium model is implemented in a {\it Lagrangian} frame of reference. This difference will be important below when imposing boundary conditions.
\begin{figure}[b]
\centering
\sidecaption[t]
\includegraphics[width=0.85\columnwidth]{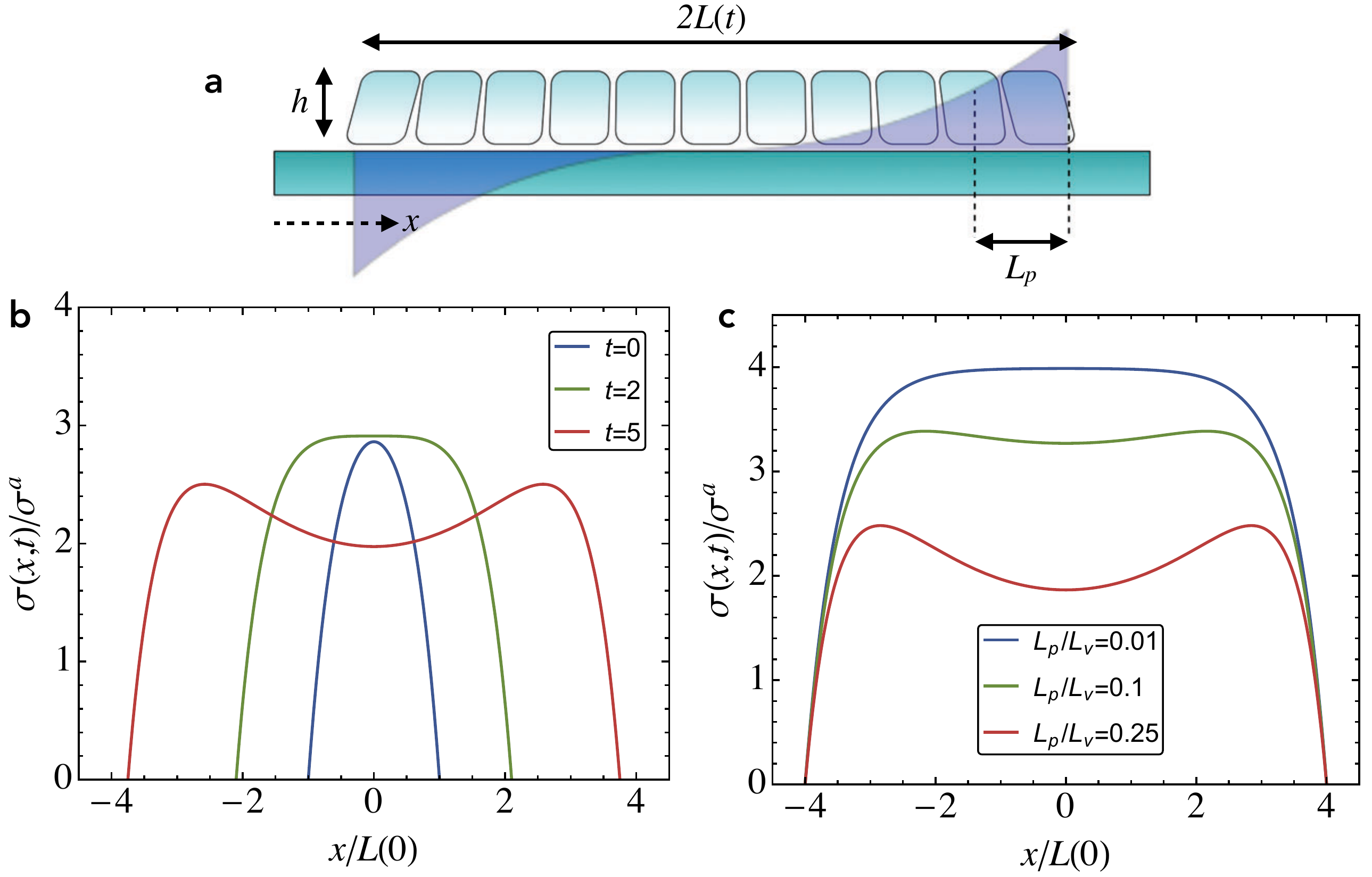}
\caption{{\bf Viscous fluid model of expanding monolayers.} (a) Schematic of an expanding epithelial monolayer of height $h$ and length $2L$, studied in Ref.~\cite{blanch2017}. The purple shaded curve represents the spatial profile of the polarization field, whose penetration depth is characterized by the length scale $L_p$. (b) Representative stress profiles of an expanding cell monolayer, predicted by the fluid model in Eq.~\eqref{eq:stress}, at different values of time with fixed $L_p/L_v=0.25$. (c) Stress profiles for different values of $L_p/L_v$ at fixed length $L=4L(0)$. Other parameters: $L_p/L(0)=0.5$, $f_0/\sigma^a=2$.}
\label{fig:10}       
\end{figure}

The case of a fluid layer of growing width $2L(t)$ was discussed in Ref.~\cite{blanch2017} (Fig.~\ref{fig:10}a). In this case intercellular stresses are purely viscous, with $\sigma^c=\eta\partial_xv_x$ and $\eta$ the shear viscosity. Assuming that cells at the boundaries are outwardly polarized to drive expansion, i.e., $p_x(-L(t))=-1$ and  $p_x(+L(t))=1$, the static polarization profile is given by
\begin{equation}
p_x(x)=\frac{\sinh(x/L_p)}{\sinh(L(t)/L_p)}\;.
\label{eq:pol_fluid}
\end{equation}
The force balance equation, Eq.~\eqref{eq:balance_1d}, can then be recast as an equation for the total stress in the fluid monolayer $\sigma_v(x)=\sigma(x)$,
\begin{equation}\label{eq:st}
\frac{1}{L_v^2} (\sigma_v - \sigma^a)=\frac{f_0}{L_p}\partial_xp_x + \partial_x^2 \sigma_v
\end{equation}
where  $L_v=\sqrt{h\eta/\zeta}$ is a viscous length scale, and $f_0=fL_p/h$ is a characteristic stress scale. We solve Eq.~\eqref{eq:st} with stress-free boundary conditions at the monolayer edge,
$\sigma_v(x=\pm L(t))=0$, where $L=L(t)$ is the growing monolayer length. The resultant stress is,
\begin{equation}\label{eq:stress}
\sigma_{v}(x)=\sigma^a\left[1-\frac{\cosh(x/L_v)}{\cosh(L/L_v)}\right]+\frac{f_0L_v^2}{L_p^2-L_v^2}\left[\frac{\cosh(x/L_p)}{\sinh(L/L_p)}-\frac{\cosh(L/L_p)\cosh(x/L_v)}{\sinh(L/L_p)\cosh(L/L_v)}\right]\;.
\end{equation}
As shown in Ref.~\cite{blanch2017} and in Fig.~\ref{fig:10}b-c, the shape of the stress profile depends on the length $L(t)$ as well as on the ratio $L_p/L_v>0$. With increasing $L$ (for fixed $L_p/L_v$) or $L_p/L_v$ (for fixed $L$), the initial stress maxima at the center of the layer disappears, and two stress peaks accumulate near the edge of the colony.

The length $L(t)$ of the expanding layer can be determined by equating the rate of change of $L(t)$ to the velocity at the leading edge, $\dot{L}=v_x(L)$. For $L(t)\gg L_p,L_v$ we find
$v_x(L)\simeq f_0L_v^2/\eta(L_p+L_v)-\sigma_aL_v/\eta$, resulting in a linear growth in time of the length of the monolayer, 
\begin{equation}
L(t)=L_0+ \frac{L_pL_v^2}{h\eta(L_p+L_v)}(f-f_c^v)t\;,
\end{equation}
provided  the pulling force $f$ exceeds a threshold value required to overcome the contractile force, $f_c^v=h\sigma^a(L_p^{-1} + L_v^{-1})$, and drive layer expansion. Note, however, that the assumption of indefinite growth in time in the absence of cell division is not realistic. Such a growth will be arrested by the requirement of volume conservation.

If the cell monolayer is modeled as an elastic continuum, then $\sigma^c=B\partial_xu_x$ where $B$ is a compressional modulus and $u_x$ the displacement field. The velocity must be identified with the rate of change of the displacement, $v_x=\partial_t u_x$. In this case the layer has a reference length $2L_0$ and an expanded length $2L(t)=2L_0+u(L_0,t)-u(-L_0,t)$. The polarization profile has the same functional form given in Eq.~\eqref{eq:pol_fluid}, but with $L(t)$ replaced by $L_0$. It is then evident that, in the absence of cell division and growth, the only steady state solution will have $v_x=0$ corresponding to the fact that the elastic layer can be stretched by outward pulling cells, but not indefinitely expanded. The stress balance equation can again be written as a closed equation for the stress ($\sigma_{el}(x)=\sigma(x)$), $h\partial_x\sigma_{el}=-fp_x(x)$, with the solution (Fig.~\ref{fig:5}b)
\begin{equation}
\sigma_{el}(x)=f_0\frac{\cosh(L_0/L_p)-\cosh(x/L_p)}{\sinh(L_0/L_p)}\;.
\end{equation}
The stress profile of the elastically stretched tissue is controlled by the single length scale $L_p$ and always shows a maximum at the midpoint of the layer. From this solution, one can immediately obtain the steady state displacement field, $u_x$, at the sample boundary, up to an undetermined constant. We eliminate this constant by assuming a symmetric deformation profile such that $u_x(0)=0$. In the limit $L_0\gg L_p$ we get
$u_x(L_0)=-u_x(-L_0)=\frac{L_0}{B}\left(f_0L_p/L_0-\sigma^a\right)$. Of course in this case the monolayer stretches only provided the pulling forces due to polarization exceed the contractile forces. There is a critical pulling force, given by $f_c^e=h\sigma^a L_0/L_p^2$.
Retaining again only leading terms in $L_p/L_0$, the total length of the expanded monolayer is then given by 
\begin{equation}
L_\infty=L_0\left[1+\frac{L_p^2}{L_0h}(f-f_c^e)\right]\;.
\end{equation}

Unlike the fluid, a purely elastic layer cannot sustain a state of steady growth. To obtain steady expansion in the case where the layer is modeled as an elastic medium it is necessary to include cell division. This can be accomplished in several ways: by allowing the reference layer length $L_0$ to grow with time; by describing cell division in terms of an extensile contribution to the active stress, such as $\sigma^{a,g}=-Rt$, where $R>0$ describes the rate of growth; or by allowing the elastic constant $B$ to vary in time. Each of these prescriptions will in general give different expansion rates for the monolayer. 
A full discussion of these cases is beyond the scope of the present review. 
In general, both viscous and elastic models have successfully reproduced the stress, velocity and deformation profiles measured in experiments. This suggests that these large scale quantities may not be terribly sensitive to the specific rheology of the monolayer. More work, however, remains to be done to fully understand the mechanisms that allow living tissues to maintain their cohesiveness, while exhibiting the fluidity necessary for motion and morphological changes, and to formulate a rheological model capable of capturing these unique properties.

\section{Conclusion}\label{sec:5}
Continuum models of multicellular mechanics have been widely successful in describing the physical forces, flow and deformation patterns that mediate collective cell migration during wound healing, tissue morphogenesis and development. These models are largely based on phenomenological approaches rooted in soft condensed matter physics, fluid dynamics and statistical mechanics~\cite{marchetti2013}. One of the key advantages of a continuum framework is that it is formulated in terms of a few coarse-grained {\it collective variables} such as density, velocity, strain and stress fields which are directly measurable in experiments. The resultant theory contains only a small number of macroscopic parameters, representing the effective mechanochemical couplings that arise from the combined effect of a number of signaling pathways at subcellular and cellular scales. 

On the other hand, continuum models are generally written down phenomenologically, leaving open the key challenge of relating  the continuum scale mechanical parameters to specific processes that control the active behavior of cells at $\mu$m and nm scales. In the absence of such a connection between subcellular and cellular or tissue scale, there are no constraints on the range of values spanned by the parameters of the continuum model. Many of the molecular pathways that mediate force generation and movement in cells are, however, intimately coupled and also sensitive to external perturbations and to the physical properties the cell's environment. It is then likely that molecular scale feedback processes may constrain the range of parameter values that are accessible at the cellular and tissue scales. As a result, all the complex dynamical phases predicted by generic continuum models may not be realizable in biological systems, as particular cells and tissues may likely operate in a narrow region of parameter space. 

Another key limitation of the continuum modeling approach lies in the assumption of fixed materials properties of tissues, which is encoded in the choice of a particular constitutive law. As discussed elsewhere~\cite{khalilgharibi2016}, tissue rheology is highly complex, and the presence of multiple relaxation times demands a rheological model capable of capturing both active solid-like and fluid-like behavior in different regimes of stress response. In this review, we focus on active elastic models of tissue mechanics~\cite{banerjee2012,mertz2012,edwards2011,kopf2013,banerjee2015,notbohm2016} which have been successful in capturing many experimentally observed cell behaviors during collective migration. These include mechanical waves~\cite{serra2012}, collective cell rotations~\cite{doxzen2013,deforet2014,notbohm2016}, traction force localization~\cite{trepat2009,mertz2012}, and mechanosensitivity to extracellular matrix properties~\cite{schwarz2013}. We also compare elasticity models against viscous fluid models of cell migration~\cite{blanch2017}, showing that macroscopic quantities and observables may not sensitive to the specific choice of tissue rheology. On the other hand, a number of mesoscopic models, such as the Vertex, Voronoi, Potts and particle-based models, have been shown to capture various aspects of tissue-scale mechanics, providing an alternate bottom-up approach that may allow us to connect molecular scale to tissue-scale properties. A systematic study of such models with an eye on developing the multi-scale mechanics of multicellular assemblies is currently lacking, and remains an open theoretical challenge at the interface of physics and biology.

Living cells are active entities, capable for instance of autonomous motion, spontaneous mechanical deformations, division and phenotypical changes. This behavior can often be modeled at the mesoscale through  internal state variables unique to living systems. In this review we have introduced two such internal state variables: the concentration of intracellular molecular active force generators and the cell polarity vector that describes the direction in which individual cells tend to move. For simplicity we have only considered the concentration of contractile units in the actomyosin cytoskeleton, that may represent, for instance, phosporylated myosins. More generally, several dynamically coupled chemical components may be needed to capture the complexity of molecular processes in the cell cytoskeleton. Multiple filaments, motors, and binding proteins compete to regulate cell homeostasis, polarization, and active force generation~\cite{suarez2016}. As more fascinating regulatory properties of the cytoskeletal machinery are being discovered, future models must attempt to incorporate such self-regulatory mechanisms controlling active cell mechanics. 

An open question is the molecular interpretation of the cell polarization.  Different interpretations have been put forward in the literature, including identifying cell polarization with the direction of lamellipodial/filopodial protrusions or with the orientation of the cell long axis associated with the alignment of actin stress fibers, although the latter provides a nematic (head-tail symmetric) degree of freedom, rather than a polar one. Regardless of its subcellular origin, cell polarity serves to dictate the direction of local motion, and is distinct from the actual direction of cell motion in a tissue that is also controlled by the forces from neighboring cells. In other words, the dynamics of the polarity vector encodes the decision-making rules for cell motility that come from the sum of mechanical and biochemical cues that an individual cell experiences from its internal as well as external environments. Given the multitude of polarity cues gathered by a cell, it remains contentious whether a single polarity state variable can fruitfully describe multiple mechanisms of active cell motility. 

Essential ingredients of the models described in this review are the feedbacks between cellular mechanics, polarized motility, and the regulatory biochemistry of actomyosin contractility. Mechanochemical coupling of cell motion, adhesion and contractility have been argued as the physical basis for tissue morphogenesis and development~\cite{howard2011}. These couplings also play an essential role in the transmission of spatial information in large cell monolayers, mediated by waves, pulses, and a {\it tug of war} between cell-cell and cell-substrate forces. Both negative and positive feedback loops are exploited by cells for robust movement and force generation. Positive feedback commonly occurs between mechanical strain and advective transport of cytoskeletal filaments and motors into regions of high contractility. These active forces compete with diffusion and elasticity to establish the spatial gradients of contractility responsible for spontaneous cell motion. On the other hand, negative feedback between mechanical strain and contractility can yield periodic cycles of tissue stiffening and fluidization, which can result in long-range propagation of mechanical waves in tissues. At present, however, these feedback mechanisms remain purely phenomenological constructs, with only qualitative support from experiments. Their direct quantification is an outstanding experimental challenge.

In the future, theorists and experimentalists will need to work together to identify and probe all the key mechanical and biochemical parameters in a single model system. Such collaborative efforts will lead the way to more quantitatively accurate models of collective cell behavior in physiology and development.

\begin{acknowledgement}
SB acknowledges support from a Strategic Fellowship at the Institute for the Physics of Living Systems at UCL, Royal Society Tata University Research Fellowship (URF\textbackslash R1\textbackslash 180187), and Human Frontiers Science Program (HFSP RGY0073/2018).
MCM was supported by the National Science Foundation at Syracuse University through awards DMR-1609208, DGE-1068780 and at KITP under Grant PHY-1748958, and by the Simons Foundation through a Targeted Grant Award No. 342354. MCM thanks the Syracuse Soft and Living Matter Program for support and the KITP for hospitality during completion of some of this work. \end{acknowledgement}

\bibliographystyle{spphys}

\begin{thebibliography}{100}
\providecommand{\url}[1]{{#1}}
\providecommand{\urlprefix}{URL }
\expandafter\ifx\csname urlstyle\endcsname\relax
  \providecommand{\doi}[1]{DOI \discretionary{}{}{}#1}\else
  \providecommand{\doi}{DOI \discretionary{}{}{}\begingroup
  \urlstyle{rm}\Url}\fi

\bibitem{friedl2009}
P.~Friedl, D.~Gilmour, Nature Reviews Molecular Cell Biology \textbf{10}(7),
  445 (2009)

\bibitem{ladoux2017}
B.~Ladoux, R.M. M{\`e}ge, Nature Reviews Molecular Cell Biology
  \textbf{18}(12), 743 (2017)

\bibitem{fenteany2000}
G.~Fenteany, P.A. Janmey, T.P. Stossel, Current Biology \textbf{10}(14), 831
  (2000)

\bibitem{begnaud2016}
S.~Begnaud, T.~Chen, D.~Delacour, R.M. M{\`e}ge, B.~Ladoux, Current Opinion in
  Cell Biology \textbf{42}, 52 (2016)

\bibitem{trepat2009}
X.~Trepat, M.R. Wasserman, T.E. Angelini, E.~Millet, D.A. Weitz, J.P. Butler,
  J.J. Fredberg, Nature Physics \textbf{5}(6), 426 (2009)

\bibitem{farooqui2005}
R.~Farooqui, G.~Fenteany, Journal of Cell Science \textbf{118}(1), 51 (2005)

\bibitem{serra2012}
X.~Serra-Picamal, V.~Conte, R.~Vincent, E.~Anon, D.T. Tambe, E.~Bazellieres,
  J.P. Butler, J.J. Fredberg, X.~Trepat, Nature Physics \textbf{8}(8), 628
  (2012)

\bibitem{doxzen2013}
K.~Doxzen, S.R.K. Vedula, M.C. Leong, H.~Hirata, N.S. Gov, A.J. Kabla,
  B.~Ladoux, C.T. Lim, Integrative Biology \textbf{5}(8), 1026 (2013)

\bibitem{deforet2014}
M.~Deforet, V.~Hakim, H.G. Yevick, G.~Duclos, P.~Silberzan, Nature
  Communications \textbf{5}, 3747 (2014)

\bibitem{lecuit2011}
T.~Lecuit, P.F. Lenne, E.~Munro, Annual Review of Cell and Developmental
  Biology \textbf{27}, 157 (2011)

\bibitem{angelini2010}
T.E. Angelini, E.~Hannezo, X.~Trepat, J.J. Fredberg, D.A. Weitz, Physical
  Review Letters \textbf{104}(16), 168104 (2010)

\bibitem{angelini2011}
T.E. Angelini, E.~Hannezo, X.~Trepat, M.~Marquez, J.J. Fredberg, D.A. Weitz,
  Proceedings of the National Academy of Sciences \textbf{108}(12), 4714 (2011)

\bibitem{discher2005}
D.E. Discher, P.~Janmey, Y.l. Wang, Science \textbf{310}(5751), 1139 (2005)

\bibitem{foty1994}
R.A. Foty, G.~Forgacs, C.M. Pfleger, M.S. Steinberg, Physical Review Letters
  \textbf{72}(14), 2298 (1994)

\bibitem{basan2009}
M.~Basan, T.~Risler, J.F. Joanny, X.~Sastre-Garau, J.~Prost, HFSP Journal
  \textbf{3}(4), 265 (2009)

\bibitem{mertz2013}
A.F. Mertz, Y.~Che, S.~Banerjee, J.M. Goldstein, K.A. Rosowski, S.F. Revilla,
  C.M. Niessen, M.C. Marchetti, E.R. Dufresne, V.~Horsley, Proceedings of the
  National Academy of Sciences \textbf{110}(3), 842 (2013)

\bibitem{maruthamuthu2011}
V.~Maruthamuthu, B.~Sabass, U.S. Schwarz, M.L. Gardel, Proceedings of the
  National Academy of Sciences \textbf{108}(12), 4708 (2011)

\bibitem{roca2017}
P.~Roca-Cusachs, V.~Conte, X.~Trepat, Nature Cell Biology \textbf{19}(7), 742
  (2017)

\bibitem{basan2013}
M.~Basan, J.~Elgeti, E.~Hannezo, W.J. Rappel, H.~Levine, Proceedings of the
  National Academy of Sciences \textbf{110}(7), 2452 (2013)

\bibitem{camley2017}
B.A. Camley, W.J. Rappel, Journal of Physics D: Applied Physics
  \textbf{50}(11), 113002 (2017)

\bibitem{honda1980}
H.~Honda, G.~Eguchi, Journal of Theoretical Biology \textbf{84}(3), 575 (1980)

\bibitem{fletcher2014}
A.G. Fletcher, M.~Osterfield, R.E. Baker, S.Y. Shvartsman, Biophysical Journal
  \textbf{106}(11), 2291 (2014)

\bibitem{li2014}
B.~Li, S.X. Sun, Biophysical Journal \textbf{107}(7), 1532 (2014)

\bibitem{bi2016}
D.~Bi, X.~Yang, M.C. Marchetti, M.L. Manning, Physical Review X \textbf{6}(2),
  021011 (2016)

\bibitem{graner1992}
F.~Graner, J.A. Glazier, Physical review letters \textbf{69}(13), 2013 (1992)

\bibitem{farhadifar2007}
R.~Farhadifar, J.C. R{\"o}per, B.~Aigouy, S.~Eaton, F.~J{\"u}licher, Current
  Biology \textbf{17}(24), 2095 (2007)

\bibitem{barton2017}
D.L. Barton, S.~Henkes, C.J. Weijer, R.~Sknepnek, PLoS Computational Biology
  \textbf{13}(6), e1005569 (2017)

\bibitem{staddon2018}
M.F. Staddon, D.~Bi, A.P. Tabatabai, V.~Ajeti, M.P. Murrell, S.~Banerjee, PLoS
  Computational Biology \textbf{14}(10), e1006502 (2018)

\bibitem{noll2017}
N.~Noll, M.~Mani, I.~Heemskerk, S.J. Streichan, B.I. Shraiman, Nature Physics
  \textbf{13}(12), 1221 (2017)

\bibitem{bi2015}
D.~Bi, J.~Lopez, J.~Schwarz, M.L. Manning, Nature Physics \textbf{11}(12), 1074
  (2015)

\bibitem{ziebert2011}
F.~Ziebert, S.~Swaminathan, I.S. Aranson, Journal of The Royal Society
  Interface p. rsif20110433 (2011)

\bibitem{prost2015}
J.~Prost, F.~J{\"u}licher, J.F. Joanny, Nature Physics \textbf{11}(2), 111
  (2015)

\bibitem{style2014}
R.W. Style, R.~Boltyanskiy, G.K. German, C.~Hyland, C.W. MacMinn, A.F. Mertz,
  L.A. Wilen, Y.~Xu, E.R. Dufresne, Soft Matter \textbf{10}(23), 4047 (2014)

\bibitem{banerjee2011}
S.~Banerjee, M.C. Marchetti, Europhysics Letters \textbf{96}(2), 28003 (2011)

\bibitem{blanch2017}
C.~Blanch-Mercader, R.~Vincent, E.~Bazelli{\`e}res, X.~Serra-Picamal,
  X.~Trepat, J.~Casademunt, Soft Matter \textbf{13}(6), 1235 (2017)

\bibitem{notbohm2016}
J.~Notbohm, S.~Banerjee, K.J. Utuje, B.~Gweon, H.~Jang, Y.~Park, J.~Shin, J.P.
  Butler, J.J. Fredberg, M.C. Marchetti, Biophysical Journal \textbf{110}(12),
  2729 (2016)

\bibitem{oakes2014}
P.W. Oakes, S.~Banerjee, M.C. Marchetti, M.L. Gardel, Biophysical Journal
  \textbf{107}(4), 825 (2014)

\bibitem{banerjee2015}
S.~Banerjee, K.J. Utuje, M.C. Marchetti, Physical Review Letters
  \textbf{114}(22), 228101 (2015)

\bibitem{mertz2012}
A.F. Mertz, S.~Banerjee, Y.~Che, G.K. German, Y.~Xu, C.~Hyland, M.C. Marchetti,
  V.~Horsley, E.R. Dufresne, Physical Review Letters \textbf{108}(19), 198101
  (2012)

\bibitem{marchetti2013}
M.C. Marchetti, J.F. Joanny, S.~Ramaswamy, T.B. Liverpool, J.~Prost, M.~Rao,
  R.A. Simha, Reviews of Modern Physics \textbf{85}(3), 1143 (2013)

\bibitem{banerjee2012}
S.~Banerjee, M.C. Marchetti, Physical Review Letters \textbf{109}(10), 108101
  (2012)

\bibitem{schwarz2013}
U.S. Schwarz, S.A. Safran, Reviews of Modern Physics \textbf{85}(3), 1327
  (2013)

\bibitem{legant2013}
W.R. Legant, C.K. Choi, J.S. Miller, L.~Shao, L.~Gao, E.~Betzig, C.S. Chen,
  Proceedings of the National Academy of Sciences \textbf{110}(3), 881 (2013)

\bibitem{bove2017}
A.~Bove, D.~Gradeci, Y.~Fujita, S.~Banerjee, G.~Charras, A.R. Lowe, Molecular
  Biology of the Cell \textbf{28}(23), 3215 (2017)

\bibitem{murray1984}
J.~Murray, G.~Oster, Journal of Mathematical Biology \textbf{19}(3), 265 (1984)

\bibitem{ranft2010}
J.~Ranft, M.~Basan, J.~Elgeti, J.F. Joanny, J.~Prost, F.~J{\"u}licher,
  Proceedings of the National Academy of Sciences \textbf{107}(49), 20863
  (2010)

\bibitem{yabunaka2017}
S.~Yabunaka, P.~Marcq, Physical Review E \textbf{96}(2), 022406 (2017)

\bibitem{khalilgharibi2016}
N.~Khalilgharibi, J.~Fouchard, P.~Recho, G.~Charras, A.~Kabla, Current Opinion
  in Cell Biology \textbf{42}, 113 (2016)

\bibitem{phillips1978}
H.~Phillips, M.~Steinberg, Journal of Cell Science \textbf{30}(1), 1 (1978)

\bibitem{guevorkian2010}
K.~Guevorkian, M.J. Colbert, M.~Durth, S.~Dufour, F.~Brochard-Wyart, Physical
  Review Letters \textbf{104}(21), 218101 (2010)

\bibitem{guillot2013}
C.~Guillot, T.~Lecuit, Science \textbf{340}(6137), 1185 (2013)

\bibitem{heisenberg2013}
C.P. Heisenberg, Y.~Bella{\"\i}che, Cell \textbf{153}(5), 948 (2013)

\bibitem{lee2011}
P.~Lee, C.W. Wolgemuth, PLoS Computational Biology \textbf{7}(3), e1002007
  (2011)

\bibitem{wayne2004}
G.~Wayne~Brodland, C.J. Wiebe, Computer methods in biomechanics and biomedical
  engineering \textbf{7}(2), 91 (2004)

\bibitem{harris2012}
A.R. Harris, L.~Peter, J.~Bellis, B.~Baum, A.J. Kabla, G.T. Charras,
  Proceedings of the National Academy of Sciences \textbf{109}(41), 16449
  (2012)

\bibitem{gonzalez2013}
D.~Gonzalez-Rodriguez, L.~Bonnemay, J.~Elgeti, S.~Dufour, D.~Cuvelier,
  F.~Brochard-Wyart, Soft Matter \textbf{9}(7), 2282 (2013)

\bibitem{khalilgharibi2018}
N.~Khalilgharibi, J.~Fouchard, N.~Asadipour, A.~Yonis, A.~Harris, P.~Mosaffa,
  Y.~Fujita, A.~Kabla, B.~Baum, J.J. Munoz, et~al., bioRxiv p. 302158 (2018)

\bibitem{tambe2011}
D.T. Tambe, C.C. Hardin, T.E. Angelini, K.~Rajendran, C.Y. Park,
  X.~Serra-Picamal, E.H. Zhou, M.H. Zaman, J.P. Butler, D.A. Weitz, et~al.,
  Nature Materials \textbf{10}(6), 469 (2011)

\bibitem{banerjee2011c}
S.~Banerjee, T.B. Liverpool, M.C. Marchetti, Europhysics Letters
  \textbf{96}(5), 58004 (2011)

\bibitem{kopf2013}
M.H. K{\"o}pf, L.M. Pismen, Soft Matter \textbf{9}(14), 3727 (2013)

\bibitem{banerjee2017}
D.S. Banerjee, A.~Munjal, T.~Lecuit, M.~Rao, Nature Communications
  \textbf{8}(1), 1121 (2017)

\bibitem{latorre2018}
E.~Latorre, S.~Kale, L.~Casares, M.~G{\'o}mez-Gonz{\'a}lez, M.~Uroz, L.~Valon,
  R.V. Nair, E.~Garreta, N.~Montserrat, A.~del Campo, B.~Ladoux, M.~Arroyo,
  X.~Trepat, Nature \textbf{563}(7730), 203 (2018)

\bibitem{murrell2015}
M.~Murrell, P.W. Oakes, M.~Lenz, M.L. Gardel, Nature Reviews Molecular Cell
  Biology \textbf{16}(8), 486 (2015)

\bibitem{banerjee2011b}
S.~Banerjee, M.C. Marchetti, Soft Matter \textbf{7}(2), 463 (2011)

\bibitem{bois2011}
J.S. Bois, F.~J{\"u}licher, S.W. Grill, Physical Review Letters
  \textbf{106}(2), 028103 (2011)

\bibitem{linsmeier2016}
I.~Linsmeier, S.~Banerjee, P.W. Oakes, W.~Jung, T.~Kim, M.P. Murrell, Nature
  Communications \textbf{7}, 12615 (2016)

\bibitem{walcott2010}
S.~Walcott, S.X. Sun, Proceedings of the National Academy of Sciences
  \textbf{107}(17), 7757 (2010)

\bibitem{brugues2014}
A.~Brugu{\'e}s, E.~Anon, V.~Conte, J.H. Veldhuis, M.~Gupta, J.~Colombelli, J.J.
  Mu{\~n}oz, G.W. Brodland, B.~Ladoux, X.~Trepat, Nature Physics
  \textbf{10}(9), 683 (2014)

\bibitem{lomakin2015}
A.J. Lomakin, K.C. Lee, S.J. Han, D.A. Bui, M.~Davidson, A.~Mogilner,
  G.~Danuser, Nature Cell Biology \textbf{17}(11), 1435 (2015)

\bibitem{suarez2016}
C.~Suarez, D.R. Kovar, Nature Reviews Molecular Cell Biology \textbf{17}(12),
  799 (2016)

\bibitem{ahmadi2006}
A.~Ahmadi, M.C. Marchetti, T.B. Liverpool, Physical Review E \textbf{74}(6),
  061913 (2006)

\bibitem{prost1995}
J.~Prost, \emph{The physics of liquid crystals}, vol.~83 (Oxford university
  press, 1995)

\bibitem{vicsek1995}
T.~Vicsek, A.~Czir{\'o}k, E.~Ben-Jacob, I.~Cohen, O.~Shochet, Physical Review
  Letters \textbf{75}(6), 1226 (1995)

\bibitem{robin2016}
F.B. Robin, J.B. Michaux, W.M. McFadden, E.M. Munro, bioRxiv p. 076356 (2016)

\bibitem{vincent2015}
R.~Vincent, E.~Bazelli{\`e}res, C.~P{\'e}rez-Gonz{\'a}lez, M.~Uroz,
  X.~Serra-Picamal, X.~Trepat, Physical Review Letters \textbf{115}(24), 248103
  (2015)

\bibitem{levayer2012}
R.~Levayer, T.~Lecuit, Trends in Cell Biology \textbf{22}(2), 61 (2012)

\bibitem{gross2017}
P.~Gross, K.V. Kumar, S.W. Grill, Annual Review of Biophysics \textbf{46}, 337
  (2017)

\bibitem{wozniak2009}
M.A. Wozniak, C.S. Chen, Nature Reviews Molecular cell biology \textbf{10}(1),
  34 (2009)

\bibitem{du2005}
O.~Du~Roure, A.~Saez, A.~Buguin, R.H. Austin, P.~Chavrier, P.~Siberzan,
  B.~Ladoux, Proceedings of the National Academy of Sciences \textbf{102}(7),
  2390 (2005)

\bibitem{edwards2011}
C.M. Edwards, U.S. Schwarz, Physical Review Letters \textbf{107}(12), 128101
  (2011)

\bibitem{ravasio2015}
A.~Ravasio, A.P. Le, T.B. Saw, V.~Tarle, H.T. Ong, C.~Bertocchi, R.M. M{\`e}ge,
  C.T. Lim, N.S. Gov, B.~Ladoux, Integrative Biology \textbf{7}(10), 1228
  (2015)

\bibitem{banerjee2013}
S.~Banerjee, M.C. Marchetti, New Journal of Physics \textbf{15}(3), 035015
  (2013)

\bibitem{ghibaudo2008}
M.~Ghibaudo, A.~Saez, L.~Trichet, A.~Xayaphoummine, J.~Browaeys, P.~Silberzan,
  A.~Buguin, B.~Ladoux, Soft Matter \textbf{4}(9), 1836 (2008)

\bibitem{chopra2011}
A.~Chopra, E.~Tabdanov, H.~Patel, P.A. Janmey, J.Y. Kresh, American Journal of
  Physiology-Heart and Circulatory Physiology \textbf{300}(4), H1252 (2011)

\bibitem{roca2008}
P.~Roca-Cusachs, J.~Alcaraz, R.~Sunyer, J.~Samitier, R.~Farr{\'e}, D.~Navajas,
  Biophysical Journal \textbf{94}(12), 4984 (2008)

\bibitem{zemel2010}
A.~Zemel, F.~Rehfeldt, A.~Brown, D.~Discher, S.~Safran, Nature Physics
  \textbf{6}(6), 468 (2010)

\bibitem{bischofs2009}
I.B. Bischofs, S.S. Schmidt, U.S. Schwarz, Physical Review Letters
  \textbf{103}(4), 048101 (2009)

\bibitem{delanoe2010}
H.~Delano{\"e}-Ayari, J.~Rieu, M.~Sano, Physical Review Letters
  \textbf{105}(24), 248103 (2010)

\bibitem{schaumann2018}
E.N. Schaumann, M.F. Staddon, M.L. Gardel, S.~Banerjee, Molecular Biology of
  the Cell \textbf{29}(23), 2835 (2018)

\bibitem{yarrow2004}
J.C. Yarrow, Z.E. Perlman, N.J. Westwood, T.J. Mitchison, BMC Biotechnology
  \textbf{4}(1), 21 (2004)

\bibitem{poujade2007}
M.~Poujade, E.~Grasland-Mongrain, A.~Hertzog, J.~Jouanneau, P.~Chavrier,
  B.~Ladoux, A.~Buguin, P.~Silberzan, Proceedings of the National Academy of
  Sciences \textbf{104}(41), 15988 (2007)

\bibitem{Petitjean2010}
L.~Petitjean, M.~Reffay, E.~Grasland-Mongrain, M.~Poujade, B.~Ladoux,
  A.~Buguin, P.~Silberzan, Biophysical Journal \textbf{98}(9), 1790 (2010)

\bibitem{thery2012}
M.~Th{\'e}ry, Nature Physics \textbf{8}(8), 583 (2012)

\bibitem{d2001}
L.A. D'Amico, M.S. Cooper, Developmental Dynamics \textbf{222}(4), 611 (2001)

\bibitem{tanner2012}
K.~Tanner, H.~Mori, R.~Mroue, A.~Bruni-Cardoso, M.J. Bissell, Proceedings of
  the National Academy of Sciences \textbf{109}(6), 1973 (2012)

\bibitem{thery2009}
M.~Th{\'e}ry, M.~Piel, Cold Spring Harbor Protocols \textbf{2009}(7), pdb
  (2009)

\bibitem{segerer2015}
F.J. Segerer, F.~Th{\"u}roff, A.P. Alberola, E.~Frey, J.O. R{\"a}dler, Physical
  Review Letters \textbf{114}(22), 228102 (2015)

\bibitem{duclos2014}
G.~Duclos, S.~Garcia, H.~Yevick, P.~Silberzan, Soft Matter \textbf{10}(14),
  2346 (2014)

\bibitem{duclos2017}
G.~Duclos, C.~Erlenk{\"a}mper, J.F. Joanny, P.~Silberzan, Nature Physics
  \textbf{13}(1), 58 (2017)

\bibitem{saw2017}
T.B. Saw, A.~Doostmohammadi, V.~Nier, L.~Kocgozlu, S.~Thampi, Y.~Toyama,
  P.~Marcq, C.T. Lim, J.M. Yeomans, B.~Ladoux, Nature \textbf{544}(7649), 212
  (2017)

\bibitem{kabla2012}
A.J. Kabla, Journal of The Royal Society Interface p. rsif20120448 (2012)

\bibitem{albert2016}
P.J. Albert, U.S. Schwarz, PLoS Computational Biology \textbf{12}(4), e1004863
  (2016)

\bibitem{wood2002}
W.~Wood, A.~Jacinto, R.~Grose, S.~Woolner, J.~Gale, C.~Wilson, P.~Martin,
  Nature Cell Biology \textbf{4}(11), 907 (2002)

\bibitem{rosenblatt2001}
J.~Rosenblatt, M.C. Raff, L.P. Cramer, Current Biology \textbf{11}(23), 1847
  (2001)

\bibitem{jacinto2001}
A.~Jacinto, A.~Martinez-Arias, P.~Martin, Nature Cell Biology \textbf{3}(5),
  E117 (2001)

\bibitem{martin1992}
P.~Martin, J.~Lewis, Nature \textbf{360}(6400), 179 (1992)

\bibitem{anon2012}
E.~Anon, X.~Serra-Picamal, P.~Hersen, N.C. Gauthier, M.P. Sheetz, X.~Trepat,
  B.~Ladoux, Proceedings of the National Academy of Sciences \textbf{109}(27),
  10891 (2012)

\bibitem{bement1993}
W.M. Bement, P.~Forscher, M.S. Mooseker, The Journal of Cell Biology
  \textbf{121}(3), 565 (1993)

\bibitem{ravasio2015b}
A.~Ravasio, I.~Cheddadi, T.~Chen, T.~Pereira, H.T. Ong, C.~Bertocchi,
  A.~Brugues, A.~Jacinto, A.J. Kabla, Y.~Toyama, et~al., Nature Communications
  \textbf{6}, 7683 (2015)

\bibitem{vedula2015}
S.R.K. Vedula, G.~Peyret, I.~Cheddadi, T.~Chen, A.~Brugu{\'e}s, H.~Hirata,
  H.~Lopez-Menendez, Y.~Toyama, L.N. De~Almeida, X.~Trepat, et~al., Nature
  Communications \textbf{6}, 6111 (2015)

\bibitem{cochet2014}
O.~Cochet-Escartin, J.~Ranft, P.~Silberzan, P.~Marcq, Biophysical Journal
  \textbf{106}(1), 65 (2014)

\bibitem{arciero2011}
J.~Arciero, Q.~Mi, M.F. Branca, D.J. Hackam, D.~Swigon, Biophysical Journal
  \textbf{100}, 535 (2011)

\bibitem{recho2016}
P.~Recho, J.~Ranft, P.~Marcq, Soft Matter \textbf{12}, 2381 (2016)

\bibitem{blanch2017-2}
C.~Blanch-Mercader, J.~Casademunt, Soft Matter \textbf{13}(38), 6913 (2017)

\bibitem{perez2018}
C.~P{\'e}rez-Gonz{\'a}lez, R.~Alert, C.~Blanch-Mercader,
  M.~G{\'o}mez-Gonz{\'a}lez, T.~Kolodziej, E.~Bazellieres, J.~Casademunt,
  X.~Trepat, Nature Physics  (2018)

\bibitem{howard2011}
J.~Howard, S.W. Grill, J.S. Bois, Nature Reviews Molecular Cell Biology
  \textbf{12}(6), 392 (2011)

\end{thebibliography}

\end{document}